\documentclass[11pt]{article}%
\usepackage{amsmath}
\usepackage{amssymb}
\usepackage{amsfonts}

\usepackage{cite}
\usepackage{graphicx}
\usepackage{float}
\usepackage{longtable}
\usepackage[utf8]{inputenc}
\usepackage{hyperref}

\usepackage{algorithmicx}
\usepackage[ruled,vlined]{algorithm2e}
\SetAlgoCaptionSeparator{ }

\usepackage{fullpage}

\newcommand{\fref}[1]{Fig.~\ref{#1}}

\newcommand{\tref}[1]{Table~\ref{#1}}
\newcommand{\sref}[1]{Section~\ref{#1}}

\providecommand{\U}[1]{\protect\rule{.1in}{.1in}}
\newtheorem{theorem}{Theorem}

\newtheorem{lemma}{Lemma}

\newtheorem{proposition}{Proposition}

\newenvironment{proof}[1][Proof]{\textbf{#1.} }{\ \rule{0.5em}{0.5em}}

\begin{document}

\title{Correlation Measure Equivalence in Dynamic Causal Structures of Quantum Gravity}
\author{Laszlo Gyongyosi\thanks{School of Electronics and Computer Science, University of Southampton, Southampton SO17 1BJ, U.K., and Department of Networked Systems and Services, Budapest University of Technology and Economics, 1117 Budapest, Hungary, and MTA-BME Information Systems Research Group, Hungarian Academy of Sciences, 1051 Budapest, Hungary.}
}

\date{}

\maketitle
\begin{abstract}
We prove an equivalence transformation between the correlation measure functions of the causally-unbiased quantum gravity space and the causally-biased standard space. The theory of quantum gravity fuses the dynamic (nonfixed) causal structure of general relativity and the quantum uncertainty of quantum mechanics. In a quantum gravity space, the events are causally nonseparable and all time bias vanishes, which makes it no possible to use the standard causally-biased entropy and the correlation measure functions. Since a corrected causally-unbiased entropy function leads to an undefined, obscure mathematical structure, in our approach the correction is made in the data representation of the causally-unbiased space. Here we prove that the standard causally-biased entropy function with a data correction can be used to identify correlations in dynamic causal structures. As a corollary, all mathematical properties of the causally-biased correlation measure functions are preserved in the causally-unbiased space. The equivalence transformation allows us to measure correlations in a quantum gravity space with the stable, well-defined mathematical background and apparatus of the causally-biased functions of quantum Shannon theory.
\end{abstract}

\section{Introduction}
\label{sec1}
The theory of quantum gravity (QG) fuses the dynamic (nonfixed) causal structure of general relativity (GR) and the quantum uncertainty of quantum mechanics (QM) \cite{ref1,ref2,ref3,ref4,ref5,ref6,ref7,ref8,ref9,ref10,ref11,ref12,ref13,ref14}. Quantum gravity information processing (QGIP) proposes a framework to perform quantum information processing \cite{ref14,ref15,ref16,ref17,ref18,ref19,ref20,ref21,ref22,ref23,ref24,ref25,ref26,ref27,ref28,ref29,ref30,ref31,ref32,ref33} and quantum computations \cite{ref33,ref34,ref35,ref36,ref37,ref38,ref39,ref40,ref41,ref42,ref43,ref44,ref45,ref46,ref47,ref48,ref49,ref50,ref51,ref52,ref53,ref54,ref55,ref50,ref56,ref57,ref58,ref59,ref60,ref61,ref62,ref63,ref64,ref65,ref66,ref67,ref68,ref69,ref70,ref71,ref72,ref73,ref74,ref75,ref76,ref77,ref78,ref78b}in a causally-unbiased space-time structure. The theory of QGIP allows us to build quantum gravity computers \cite{ref3} that are not just equipped with the power of quantum computers \cite{ref79,ref80,ref81,ref82,ref83,ref84,ref85,ref86} but also operating on a dynamic causal structure \cite{ref4}.

While in our causally-biased standard space-time structure the background time has an interpretable meaning, in a quantum gravity environment, the processes and events are causally nonseparable and the background time and the time steps have no interpretable meaning. 

In particular, to handle the entropy function and correlation measures in a causally-unbiased structure, the entropy function of the standard causally-biased space has been redefined \cite{ref5}. The definition of causally-unbiased entropy has in fact resulted from a correction to the causally-biased definition of entropy function. Precisely, this correction is justified by the fact that in a causally-unbiased structure, the immediate past has no interpretable meaning and has been demonstrated through the so-called causaloid framework \cite{ref5}. On the other hand, this correction in the definition of causally-unbiased entropy function has raised several questions regarding the mathematical structure and properties of the causally-unbiased entropic function. Particularly, the mathematical background of the corrected entropy function approach is undefined and obscure, which does not allow us to answer the question of which mathematical properties of causally-biased entropy hold for causally-unbiased entropy. The corrected entropy function does not provide a stable ground to construct further well-defined correlation measure functions based on it. 

Here, we propose a different approach to identify entropic measures in the causally-unbiased quantum gravity space such that all mathematical properties of the causally-biased entropy function are completely preserved. In comparison with the correction of the entropy function, in our approach, the correction is made in the data representation of the causally-unbiased space. As a convenient result, the standard causally-biased entropy function can be used to measure probabilistic correlations in the quantum gravity space such that all the mathematical properties of the causally-biased functions are preserved. The results are also extended to the corresponding correlation measure functions of quantum Shannon theory.

We prove an equivalence transformation between the correlation measure functions of the causally-unbiased quantum gravity space and the causally-biased standard space. Based on the data correction of the equivalence transformation, all mathematical properties of the standard correlation measures are preserved in the causally-unbiased quantum gravity space. Using Minkowski diagrams and the framework of Lorentz transformations \cite{ref6,ref7,ref8}, we represent the data correction and the equivalence transformation between the quantum gravity space and the standard space. The information propagation between distant parties is modeled via abstract light pulses, while the events of the Minkowski space represents correlated data between the parties. 

In the causally-unbiased quantum gravity space, all time information of all events vanishes, which does not allow the interpretation of any causal connection between the events. In particular, we show that our equivalence transformation can project the events of the causally-unbiased quantum gravity space onto the causally-biased standard space, with interpretable time bias and causal connection between them.

The novel contributions of our paper are as follows:
\begin{enumerate}
\item  We prove a mathematical equivalence transformation between the correlation measure functions of the causally-unbiased quantum gravity space and the causally-biased standard space. 

\item  We prove that the standard causally-biased entropy function with a data correction can be used to identify correlations in dynamic causal structures. 

\item  We show that all mathematical properties of the causally-biased correlation measure functions are preserved in the causally-unbiased space. 

\item  We prove that it is possible to achieve a correction in the data representation of the causally-unbiased space. 

\item  We prove that the equivalence transformation allows us to measure correlations in a quantum gravity space with the stable, well-defined mathematical structure.
\end{enumerate}

This paper is organized as follows. In \sref{sec2}, some preliminary findings are summarized. \sref{sec3} discusses the data representation correction and provides the proof equivalence transformation. \sref{sec4} proposes the correlation measure functions for a quantum gravity space. Finally, \sref{sec5} concludes the results. Supplementary information is included in the Appendix.

\section{Preliminaries}
\label{sec2}
\subsection{Terms and Notations }
In this section, we summarize the basic terms of the manuscript based on the notations of \cite{ref5}.

\subsubsection{Data}
A piece of data $d$ results from a measurement ${\rm {\mathcal M}}$, which is identified in a space-time structure via 
\begin{equation} \label{1)} 
d=\left(x,\varphi _{x} ,y_{x} \right),                                                  
\end{equation} 
where $x$ is a space-time coordination (elementary region of the space) where the measurement is made in the space, $\varphi _{x} $ refers to the information pertinent to a choice of ${\rm {\mathcal M}}$, and $y_{x} $ denotes the outcome of ${\rm {\mathcal M}}$.

\subsubsection{Measurement Information}
The elementary region $R_{x} $ is the set of all possible data $d_{i} $, where the space-time information is $x_{i} =x$ for $\forall i$ (measurement information for elementary region $x$),
\begin{equation} \label{2)} 
\begin{split}
   {{R}_{x}}&=\bigcup\limits_{\forall {{x}_{i}}=x}{{{d}_{i}}} \\ 
 & =\bigcup\limits_{\forall {{x}_{i}}=x}{\left( {{x}_{i}},{{\varphi }_{{{x}_{i}}}},{{y}_{{{x}_{i}}}} \right)}.  
\end{split}
\end{equation} 
A composite region ${\rm {\mathcal O}}_{C} $ is a set of $i$ elementary regions allowing the definition of measurement information $R_{C} $ for the elementary regions of ${\rm {\mathcal O}}_{C} $ as
\begin{equation} \label{3)} 
R_{C} =\bigcup _{x\in {\rm {\mathcal O}}_{C} }R_{x}  .                                                  
\end{equation} 
Then, let $V$ refer to a set of all possible $R_{C} $ measurement information from all elementary regions $x$ as 
\begin{equation} \label{4)} 
V=\bigcup _{\forall x}R_{x}  ,                                                   
\end{equation} 
where $R_{x} $ is the measurement information for an elementary region $x$.

\subsubsection{Procedure}
Let $F_{x} $ refer to the procedure in region $x$, which identifies the set of all distinct choices of measurement for the elementary region $x$ as 
\begin{equation} \label{5)} 
F_{x} =\bigcup _{\forall \varphi _{x} ,y_{x} }\left(x,\varphi _{x} ,y_{x} \right) .                                          
\end{equation} 
Using ${\rm {\mathcal O}}_{C} $, $F_{C} $ can be identified as
\begin{equation} \label{6)} 
F_{C} =\bigcup _{x\in {\rm {\mathcal O}}_{C} }F_{x}  .                                                 
\end{equation}

\subsubsection{Measurement}
Let $Y_{x} $ refer to the outcome set for a region $x$, which identifies the set of all distinct outcomes of a measurement ${\rm {\mathcal M}}$ for $x$, 
\begin{equation} \label{7)} 
Y_{x} =\bigcup _{\forall y_{x} }\left(x,\varphi _{x} ,y_{x} \right) ,                                           
\end{equation} 
where 
\begin{equation} \label{8)} 
Y_{x} \subseteq F_{x} \subseteq R_{x} .                                               
\end{equation} 
For a composite region ${\rm {\mathcal O}}_{C} $, $Y_{C} $ is evaluated as
\begin{equation} \label{9)} 
Y_{C} =\bigcup _{x\in {\rm {\mathcal O}}_{C} }Y_{x}  ,                                                
\end{equation} 
where
\begin{equation} \label{10)} 
Y_{C} \subseteq F_{C} \subseteq R_{C} .                                             
\end{equation}

\subsubsection{Reference Region}
The reference region $x_{A} $ is an arbitrary region in the space, and defines $F_{A} $, $Y_{A} $ as
\begin{equation} \label{11)} 
F_{A} =\bigcup _{\forall \varphi _{x_{A} } ,y_{x_{A} } }\left(x_{A} ,\varphi _{x_{A} } ,y_{x_{A} } \right) , 
\end{equation} 
\begin{equation} \label{12)} 
Y_{A} =\bigcup _{\forall y_{x_{A} } }\left(x_{A} ,\varphi _{x_{A} } ,y_{x_{A} } \right).  
\end{equation}

\subsubsection{Region of Interest }
The region of interest $x_{B} $ is a region of the space with respect to the reference region $A$ and defines $F_{B} $, $Y_{B} $ as
\begin{equation} \label{13)} 
F_{B} =\bigcup _{\forall \varphi _{x_{B} } ,y_{x_{B} } }\left(x_{B} ,\varphi _{x_{B} } ,y_{x_{B} } \right) , 
\end{equation} 
\begin{equation} \label{14)} 
Y_{B} =\bigcup _{\forall y_{x_{B} } }\left(x_{B} ,\varphi _{x_{B} } ,y_{x_{B} } \right).  
\end{equation} 

\subsubsection{Event }
For an elementary region $x$, an event $E$ is identified via a measurement-procedure pair
\begin{equation}\label{15)}
\begin{split}
   E&:\left\{ {{F}_{x}},{{Y}_{x}} \right\} \\ 
 & =\left\{ \bigcup\limits_{\forall {{\varphi }_{x}},{{y}_{x}}}{\left( x,{{\varphi }_{x}},{{y}_{x}} \right)},\bigcup\limits_{\forall {{y}_{x}}}{\left( x,{{\varphi }_{x}},{{y}_{x}} \right)} \right\}.  
\end{split}
\end{equation}

\subsubsection{Reference Event }
The reference event $A$ is defined by the reference elementary region $x_{A} $ via the outcome-measurement pair $F_{A} ,Y_{A} $, 
\begin{equation} \label{16)} 
A:\left\{F_{A} ,Y_{A} \right\}.                                              
\end{equation}

\subsubsection{Event of Interest }
The event of interest $B$ is defined via the region of interest $x_{B} $ as 
\begin{equation} \label{17)} 
B:\left\{F_{B} ,Y_{B}^{i} \right\},                                              
\end{equation} 
where $Y_{B}^{i} $ denotes a set of outcomes corresponding to $F_{B} $. 

\subsubsection{Entropy}
Let ${\rm {\mathcal S}}$ refer to the causally-biased standard space-time structure, and ${\rm {\mathcal Q}{\mathcal G}}$ to the quantum gravity (causally-unbiased) spacetime. 

\paragraph{Standard Space}
In the causally-biased ${\rm {\mathcal S}}$-spacetime, the reference event $A:\left\{F_{A} ,Y_{A} \right\}\in {\rm {\mathcal S}}$ belongs to an immediate past region, $t_{A} <t_{B} $, and the event of interest is defined via $B:\left\{F_{B} ,Y_{B}^{i} \right\}\in {\rm {\mathcal S}}$. The $p_{i}^{{\rm {\mathcal S}}} \left(\cdot \right)$ probability function in the ${\rm {\mathcal S}}$-spacetime is yielded as
\begin{equation} \label{ZEqnNum772140} 
p_{i}^{{\rm {\mathcal S}}} \left(B\right)=\Pr \left(\left. Y_{B}^{i} \right|F_{B} ,D_{A} \right),                                       
\end{equation} 
where $D_{A} $ refers to sufficient data from the past space-time region $A\in {\rm {\mathcal S}}$. 

Specifically, from \eqref{ZEqnNum772140}, the $H^{{\rm {\mathcal S}}} \left(B\right)$ entropy function in the ${\rm {\mathcal S}}$-spacetime is yielded as 
\begin{equation} \label{19)}
\begin{split}
   {{H}^{\mathcal{S}}}\left( B \right)&=-\sum\limits_{i}{p_{i}^{\mathcal{S}}}{{\log }_{2}}p_{i}^{\mathcal{S}} \\ 
 & =-\sum\limits_{i}{\Pr \left( \left. Y_{B}^{i} \right|{{F}_{B}},{{D}_{A}} \right)}{{\log }_{2}}\Pr \left( \left. Y_{B}^{i} \right|{{F}_{B}},{{D}_{A}} \right).  
\end{split}
\end{equation}

\paragraph{Quantum Gravity Space}
In the ${\rm {\mathcal Q}{\mathcal G}}$-spacetime all time bias vanishes, therefore the $A:\left\{F'_{A} ,Y'_{A} \right\}\in {\rm {\mathcal Q}{\mathcal G}}$ reference system is from an arbitrary region of the ${\rm {\mathcal Q}{\mathcal G}}$-spacetime. Let $B:\left\{F'_{B} ,{Y'_{B}}^{i} \right\}\in {\rm {\mathcal Q}{\mathcal G}}$ be the event of interest in the ${\rm {\mathcal Q}{\mathcal G}}$-spacetime, and let $H^{{\rm {\mathcal Q}{\mathcal G}}} \left(\cdot \right)$ refer to the ${\rm {\mathcal Q}{\mathcal G}}$-spacetime entropy function, respectively. 

Then, the $H^{{\rm {\mathcal Q}{\mathcal G}}} \left(B\right)$ entropy relative to event $A\in {\rm {\mathcal Q}{\mathcal G}}$ is as 
\begin{equation} \label{20)} 
H^{{\rm {\mathcal Q}{\mathcal G}}} \left(B\right)=-\sum _{i}p_{i}^{{\rm {\mathcal Q}{\mathcal G}}}  \log _{2} p_{i}^{{\rm {\mathcal Q}{\mathcal G}}} ,                                    
\end{equation} 
where $p_{i}^{{\rm {\mathcal Q}{\mathcal G}}} $ is 
\begin{equation} \label{21)} 
p_{i}^{{\rm {\mathcal Q}{\mathcal G}}} \left(B\right)=\Pr \left(\left. {Y'_{B}}^{i} \right|Y'_{A} ,F'_{A} ,F'_{B} \right),                                    
\end{equation} 
from which  
\begin{equation} \label{22)} 
H^{{\rm {\mathcal Q}{\mathcal G}}} \left(B\right)=-\sum _{i}\Pr \left(\left. {Y'_{B}}^{i} \right|Y'_{A} ,F'_{A} ,F'_{B} \right) \log _{2} \Pr \left(\left. {Y'_{B}}^{i} \right|Y'_{A} ,F'_{A} ,F'_{B} \right).                 
\end{equation}

\subsection{Minkowski Diagram}
The $M$ Minkowski diagram is a space-time diagram that represents events and sequences of events in the space-time \cite{ref6,ref7,ref8}. 

In particular, a $d=2$ dimensional $M$-diagram has two coordinates: the $x$ coordinate identifies the location information of events, while the $ct$-axis characterizes the time information multiplied by $c$. In our representation, $c$ is a constant that refers to the speed of light in the ${\rm {\mathcal Q}{\mathcal G}}$-spacetime (e.g., speed of information propagation in the ${\rm {\mathcal Q}{\mathcal G}}$-spacetime). 

Particularly, an event $E$ in the a space-time ${\rm {\mathcal S}}$ (frame) is identified via $E=\left(x_{E} ,ct_{E} \right)$. Let $x,y,z,t$ be the coordinates of the four-dimensional space-time ${\rm {\mathcal S}}$. 

Then let 
\begin{equation} \label{23)} 
\left(\vec{r},ct\right)\equiv \left(x,y,z,ct\right).                                              
\end{equation} 
Under a Lorentz transformation for frames ${\rm {\mathcal S}}\to {\rm {\mathcal S}}'$,
\begin{equation} \label{24)} 
s^{2} =c^{2} t^{2} -\vec{r}^{2} , 
\end{equation} 
\begin{equation} \label{ZEqnNum333895} 
\vec{r}^{2} =x^{2} +y^{2} +z^{2} , 
\end{equation} 
one finds
\begin{equation} \label{26)} 
s^{2} =s'^{2} ,                                                      
\end{equation} 
from which the separation of the space-time via $s^{2} $ is as
\begin{equation} \label{27)} 
s^{2} =\left\{\begin{array}{l} {>0: \text{time-like separation}} \\ {=0:\text{light-like separation}} \\ {<0: \text{space-like separation}} \end{array}\right. . 
\end{equation} 
For $s^{2} \ge 0$, causal connection is possible between the events. On the other hand, for $s^{2} <0$, the spatial separation of the events is greater than the distance light can travel between the events. A light signal emitted from the origin $\left(x=0,ct=0\right)$ defines a light cone; the time-like separated events are within the light cone, the light-like separated events are on the cone, while the space-like separated events are outside the light cone. 

Assuming events $E_{1} =\left(\vec{r}_{E_{1} } ,ct_{E_{1} } \right)$ and $E_{2} =\left(\vec{r}_{E_{2} } ,ct_{E_{2} } \right)$, $s_{E_{1} E_{2} }^{2} $ is also a Lorentz invariant, 
\begin{equation} \label{28)} 
s_{E_{1} E_{2} }^{2} =c^{2} \left(t_{E_{1} } -t_{E_{2} } \right)^{2} -\left|\vec{r}_{E_{1} } -\vec{r}_{E_{2} } \right|^{2} .                                  
\end{equation}

\subsubsection{Space-like Separation}

If $E_{1} =\left(\vec{r}_{E_{1} } ,ct_{E_{1} } \right)$ and $E_{2} =\left(\vec{r}_{E_{2} } ,ct_{E_{2} } \right)$ are space-like separated, the events cannot be connected by a light signal \cite{ref6,ref7}; thus, there is no causal connection between $E_{1} $ and $E_{2} $, $s_{E_{1} E_{2} }^{2} <0$ and 
\begin{equation} \label{29)} 
x_{E_{1} } -x_{E_{2} } >c\left(t_{E_{1} } -t_{E_{2} } \right).                                        
\end{equation} 
In this case, it is possible to find a Lorentz transformation to a frame ${\rm {\mathcal S}}'$, where $E_{1} =\left(x'_{E_{1} } ,t'_{E_{1} } \right)$ and $E_{2} =\left(x'_{E_{2} } ,t'_{E_{2} } \right)$ are happening simultaneously, $t'_{E_{1} } =t'_{E_{2} } $, as
\begin{equation} \label{30)} 
c\left(t'_{E_{1} } -t'_{E_{2} } \right)=\kappa \left(c\left(t_{E_{1} } -t_{E_{2} } \right)-\frac{\Omega }{c} \left(x_{E_{1} } -x_{E_{2} } \right)\right),                            
\end{equation} 
where $0\le \Omega <c$, 
\begin{equation} \label{31)} 
0\le {\Omega \mathord{\left/ {\vphantom {\Omega  c}} \right. \kern-\nulldelimiterspace} c} <1,                                                 
\end{equation} 
while 
\begin{equation} \label{32)} 
\kappa ={1\mathord{\left/ {\vphantom {1 \sqrt{\left(1-\left({\Omega \mathord{\left/ {\vphantom {\Omega  c}} \right. \kern-\nulldelimiterspace} c} \right)^{2} \right)} }} \right. \kern-\nulldelimiterspace} \sqrt{\left(1-\left({\Omega \mathord{\left/ {\vphantom {\Omega  c}} \right. \kern-\nulldelimiterspace} c} \right)^{2} \right)} }  
\end{equation} 
is a scaling parameter; thus, 
\begin{equation} \label{33)} 
\Omega =c{\textstyle\frac{c\left(t_{E_{1} } -t_{E_{2} } \right)}{x_{E_{1} } -x_{E_{2} } }} <c.                                             
\end{equation} 
Specifically, for space-like-separated events there exists a Lorentz transformation such that the causal order of the events is reversed.

\subsubsection{Time-like Separation}

If $E_{1} =\left(\vec{r}_{E_{1} } ,ct_{E_{1} } \right)$ and $E_{2} =\left(\vec{r}_{E_{2} } ,ct_{E_{2} } \right)$ are time-like-separated, the events can be connected by a light signal, $s_{E_{1} E_{2} }^{2} >0$ \cite{ref6,ref7}; thus, there can be a causal connection between $E_{1} $ and $E_{2} $, 
\begin{equation} \label{34)} 
x_{E_{1} } -x_{E_{2} } <c\left(t_{E_{1} } -t_{E_{2} } \right).                                          
\end{equation} 
Particularly, in this case there exists no Lorentz transformation to ${\rm {\mathcal S}}'$, where $E_{1} =\left(x'_{E_{1} } ,t'_{E_{1} } \right)$ and $E_{2} =\left(x'_{E_{2} } ,t'_{E_{2} } \right)$ are happening simultaneously since it would require faster-than-light speed (e.g., faster-than-light information flowing), 
\begin{equation} \label{35)} 
\Omega =c{\textstyle\frac{c\left(t_{E_{1} } -t_{E_{2} } \right)}{x_{E_{1} } -x_{E_{2} } }} >c,                                              
\end{equation} 
from which follows that the causal direction cannot be changed. 

On the other hand, as 
\begin{equation} \label{36)} 
x'_{E_{1} } -x'_{E_{2} } =\kappa \left(\left(x_{E_{1} } -x_{E_{2} } \right)-\Omega \left(t_{E_{1} } -t_{E_{2} } \right)\right), 
\end{equation} 
there exists the frame ${\rm {\mathcal S}}'$ where $E_{1} =\left(x'_{E_{1} } ,t'_{E_{1} } \right)$ and $E_{2} =\left(x'_{E_{2} } ,t'_{E_{2} } \right)$ are happening at the same place, i.e., $x'_{E_{1} } =x'_{E_{2} } $.

\subsubsection{Light-like Separation}

For the light-like-separated $E_{1} =\left(\vec{r}_{E_{1} } ,ct_{E_{1} } \right)$ and $E_{2} =\left(\vec{r}_{E_{2} } ,ct_{E_{2} } \right)$, the events are connected via a lightline; therefore, a causal connection is possible, in this case $s_{E_{1} E_{2} }^{2} =0$, 
\begin{equation} \label{37)} 
x_{E_{1} } -x_{E_{2} } =c\left(t_{E_{1} } -t_{E_{2} } \right) 
\end{equation} 
\begin{equation} \label{38)} 
\Omega =c{\textstyle\frac{c\left(t_{E_{1} } -t_{E_{2} } \right)}{x_{E_{1} } -x_{E_{2} } }} =c;                                               
\end{equation} 
thus,
\begin{equation} \label{39)} 
c^{2} \left(t'_{E_{1} } -t'_{E_{2} } \right)=\kappa \left(c\left(t_{E_{1} } -t_{E_{2} } \right)-\left(x_{E_{1} } -x_{E_{2} } \right)\right).                             
\end{equation}

\subsubsection{Scaling}

Let the frames ${\rm {\mathcal S}}$, ${\rm {\mathcal S}}'$ with $0\le {\Omega \mathord{\left/ {\vphantom {\Omega  c}} \right. \kern-\nulldelimiterspace} c} <1$, and let $t=t'=0$; then $ct'$-axis of ${\rm {\mathcal S}}'$ is determined via \cite{ref6,ref7} 
\begin{equation} \label{40)} 
x'=0=\kappa \left(x-\Omega t\right),                                           
\end{equation} 
or equivalently,
\begin{equation} \label{41)} 
ct={\textstyle\frac{1}{{\Omega \mathord{\left/ {\vphantom {\Omega  c}} \right. \kern-\nulldelimiterspace} c} }} x.                                                     
\end{equation} 
The $x'$-axis of ${\rm {\mathcal S}}'$ is defined via 
\begin{equation} \label{42)} 
t'=0=\kappa \left(t-{\textstyle\frac{\Omega }{c^{2} x}} \right),                                                
\end{equation} 
or equivalently, 
\begin{equation} \label{43)} 
ct={\textstyle\frac{\Omega }{{cx\mathord{\left/ {\vphantom {cx c}} \right. \kern-\nulldelimiterspace} c} }} .                                                      
\end{equation} 
The Lorentz transformation ${\rm {\mathcal S}}\to {\rm {\mathcal S}}'$ changes the scale of the axes. 

Defining a length unit by $\bar{s}^{2} =-1$ leads to a hyperbola with all length $\bar{s}^{2} =-1$ for all points, as
\begin{equation} \label{44)} 
x^{2} =\left(ct\right)^{2} +1,                                                 
\end{equation} 
such that $t=0$ cuts $x$-axis at $x=1$. 

Since $s^{2} $ is a Lorentz invariant \cite{ref7,ref8}, 
\begin{equation} \label{45)} 
x'^{2} =\left(ct'\right)^{2} +1.                                                 
\end{equation}

\subsubsection{Causality }
Let  $O=\left(x_{O} =0,ct_{O} =0\right)$ and $A=\left(x_{A} ,ct_{A} \right)$ be an event pair in the causally-biased ${\rm {\mathcal S}}$-spacetime, such that $ct_{O} <ct_{A} $. The line between events $O$ and $A$ has a gradient 
\begin{equation} \label{46)}
\begin{split}
   \partial &=\tfrac{\Delta x}{\Delta ct} \\ 
 & =\tfrac{\left| {{x}_{A}}-{{x}_{O}} \right|}{\left| c{{t}_{A}}-c{{t}_{O}} \right|},  
\end{split}
\end{equation}
which characterizes the causal connection between $O$ and $A$ as 
\begin{equation} \label{47)} 
\partial =\left\{\begin{array}{l} {\le 1:O{\rm \; and\; }A \text{ causally-connected}} \\ {>1:O{\rm \; and\; }A \text{ causally not connected.}} \end{array}\right.  
\end{equation} 
In particular, for $\partial =1$, the events can be connected via a lightline, which, from a communication theory perspective, refers to the information flow between $O$ and $A$. 

\begin{center}
\begin{figure*}[htbp]
\begin{center}
\includegraphics[angle = 0,width=1\linewidth]{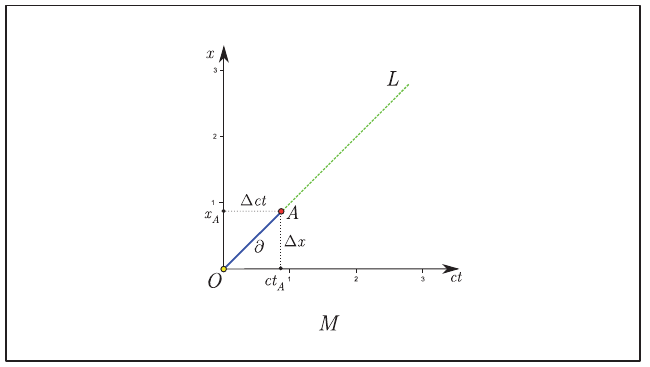}
\caption{Causally connected events $O$ and $A$, with gradient $\partial =1$.} 
 \label{fig1}
 \end{center}
\end{figure*}
\end{center}

\fref{fig1} illustrates the $M$-diagram, $d=2$, with causally connected events $O$ and $A$ in the  ${\rm {\mathcal S}}$-spacetime. The observer $O$ is in the origin, the lightline $OL$ refers to the light pulse emitted from $O$. The $L$ light ray travels along the $x$-axis toward the positive values of $x$.

\subsubsection{Local Systems, Event Coordinator, Flags}
In the further parts, events $A$ and $B$ refer to correlated local systems of two parties, Alice and Bob. 

The information flowing is modeled via flags (lightlines) $L$. In a causally-biased structure, an $O_{ec} $ event coordinator sends flags to the local parties: flag $L_{1} $ to $A$ and flag $L_{2} $ to $B$. Alice and Bob reveal their local systems as they receive flags $L_{1} $, $L_{2} $ from $O_{ec} $. In a causally-biased structure, $O_{ec} $ is causally connected to $A$ via $L_{1} $, and to $B$ via $L_{2} $.

In a causally-unbiased structure, the background time has no interpretable meaning; therefore, the flags of an $O'_{ec} $ cannot coordinate the revealing process of the local systems $A$, $B$. 

The $O_{ec} $, $O'_{ec} $ event coordinators are in the origin of the ${\rm {\mathcal S}}$-spacetime and the ${\rm {\mathcal Q}{\mathcal G}}$-spacetime, respectively, 
\begin{equation} \label{48)} 
O_{ec} =\left(x_{O_{ec} } =0,ct_{O_{ec} } =0\right) 
\end{equation} 
and 
\begin{equation} \label{49)} 
O'_{ec} =\left(x'_{O_{ec} } =0,ct'_{O_{ec} } =0\right).                                         
\end{equation} 
Modeling the information propagation via abstract light pulses in the ${\rm {\mathcal Q}{\mathcal G}}$-spacetime also allows us to build up an arbitrary quantum gravity function that constitutes a program, ${\rm {\mathcal F}}$, as
\begin{equation} \label{50)} 
{\rm {\mathcal F}}\left(x',n\right)={\rm i},                                                 
\end{equation} 
where $x'\in {\rm {\mathcal Q}{\mathcal G}}$ is an elementary region of the ${\rm {\mathcal Q}{\mathcal G}}$-spacetime and $n$ labels the corresponding flag beam, while ${\rm i}$ refers to the input such that for ${\rm i}=1$, a light pulse is emitted \cite{ref3}. Specifically, the ${\rm {\mathcal Q}{\mathcal G}}$-spacetime is normalized throughout; therefore, it is modeled via a not curved background. 

\section{Correlation Measure Equivalence}
\label{sec3}
\subsection{Correction of Data Representation}
\begin{proposition}
An $x'$ elementary region of the ${\rm {\mathcal Q}{\mathcal G}}$-spacetime can be transformed onto the ${\rm {\mathcal S}}$-spacetime via $x=\eta x'$, where $\eta ={1\mathord{\left/ {\vphantom {1 \sqrt{\left(1-\beta ^{2} \right)} }} \right. \kern-\nulldelimiterspace} \sqrt{\left(1-\beta ^{2} \right)} } $, where $\beta $, $0\le \beta <1$ is the gravity strength of the ${\rm {\mathcal Q}{\mathcal G}}$-spacetime relative to the ${\rm {\mathcal S}}$-spacetime gravity strength $\beta _{{\rm {\mathcal S}}} =0$. 
\end{proposition}
\begin{proof}
The causally-biased standard ${\rm {\mathcal S}}$-spacetime provides a reference frame with axes $x$ and $ct$. The causally-unbiased ${\rm {\mathcal Q}{\mathcal G}}$-spacetime is defined via the scaled axes $x'$ and $ct'$, with origin
\begin{equation} \label{51)} 
O'=\left(x=0,ct\right).                                              
\end{equation} 
The angle difference of the axes $\left\{x,x'\right\}$, $\left\{ct,ct'\right\}$ is $\theta $, where 
\begin{equation} \label{52)} 
0^{\circ } \le \theta <45^{\circ } .                                                
\end{equation} 
The $\theta $ angle is determined via 
\begin{equation} \label{ZEqnNum470950} 
\beta =\tan \left(\theta \right),                                                 
\end{equation} 
where $\beta $ refers to the strength of the gravity of the ${\rm {\mathcal Q}{\mathcal G}}$-spacetime relative to the ${\rm {\mathcal S}}$-spacetime gravity strength, 
\begin{equation} \label{54)} 
0\le \beta <1,                                                  
\end{equation} 
while the strength parameter of the ${\rm {\mathcal S}}$-spacetime is fixed to zero, 
\begin{equation} \label{55)} 
\beta _{{\rm {\mathcal S}}} =0.                                                     
\end{equation} 
Particularly, since in the ${\rm {\mathcal Q}{\mathcal G}}$-spacetime the background time has no interpretable meaning, $ct'=0$; that is, all events of the ${\rm {\mathcal S}}$-spacetime are projected onto the $x'$-axis of the ${\rm {\mathcal Q}{\mathcal G}}$-spacetime frame. 

Let $E$ refer to a reference event in the ${\rm {\mathcal S}}$-spacetime, 
\begin{equation} \label{56)} 
E=\left(x_{E} ,ct_{E} \right)\in {\rm {\mathcal S}}.                                            
\end{equation} 
In the ${\rm {\mathcal Q}{\mathcal G}}$-spacetime, event $E$ is identified with a location parameter, $x'_{E} \ne x_{E} ,$ and a vanishing time coefficient $ct'_{E} =0$, 
\begin{equation} \label{57)} 
E'=\left(x'_{E} ,ct'_{E} =0\right)\in {\rm {\mathcal Q}{\mathcal G}}.                                       
\end{equation} 
Without loss of generality, at a given ${\rm {\mathcal Q}{\mathcal G}}$-spacetime strength $\beta $, the $x'_{E} $ elementary region in the ${\rm {\mathcal Q}{\mathcal G}}$-spacetime in the function of $x'_{E} $ of ${\rm {\mathcal S}}$ can be expressed as
\begin{equation} \label{58)} 
x'_{E} ={\textstyle\frac{x_{E} }{\eta }} ,                                                   
\end{equation} 
where 
\begin{equation} \label{59)} 
\eta ={\textstyle\frac{1}{\sqrt{1-\beta ^{2} } }} ,                                                  
\end{equation} 
while
\begin{equation} \label{60)} 
ct'_{E} =0.                                                    
\end{equation} 
Let $l_{scale}^{{\rm {\mathcal Q}{\mathcal G}}} $ be the unit scale of the axes of $\left\{x',ct'\right\}$ of the ${\rm {\mathcal Q}{\mathcal G}}$-spacetime and $l_{scale}^{{\rm {\mathcal S}}} $ the unit scale of the axes of $\left\{x,ct\right\}$ of the ${\rm {\mathcal S}}$-spacetime. 

In particular, the axes of the ${\rm {\mathcal Q}{\mathcal G}}$-spacetime with respect to the ${\rm {\mathcal S}}$-spacetime are scaled as
\begin{equation} \label{61)} 
l_{scale}^{{\rm {\mathcal Q}{\mathcal G}}} =\gamma l_{scale}^{{\rm {\mathcal S}}} ,                                               
\end{equation} 
where $\gamma $ is the scaling parameter evaluated as
\begin{equation} \label{62)} 
\gamma =\sqrt{{\textstyle\frac{1+\beta ^{2} }{1-\beta ^{2} }} } .                                                   
\end{equation} 
\end{proof}

\begin{center}
\begin{figure*}[htbp]
\begin{center}
\includegraphics[angle = 0,width=1\linewidth]{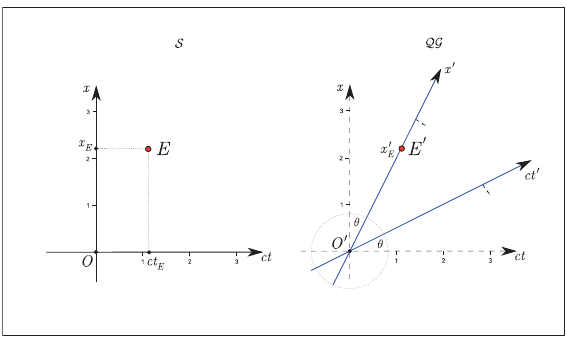}
\caption{The causally-biased standard ${\rm {\mathcal S}}$-spacetime provides a reference frame with axes $x$ and $ct$. The causally-unbiased ${\rm {\mathcal Q}{\mathcal G}}$-spacetime is defined via the scaled axes $x'$ and $ct'$ with origin  $O'=\left(x=0,ct=0\right)$. The difference of the frame axes is $0^{\circ } \le \theta <45^{\circ } $, where $\theta $ is determined via the $\beta $ strength parameter of the quantum gravity space, $0\le \beta <1$, as $\beta =\tan \left(\theta \right)$ ($\beta _{{\rm {\mathcal S}}} =0$ for the ${\rm {\mathcal S}}$-spacetime). A given ${\rm {\mathcal Q}{\mathcal G}}$-spacetime event, $E'=\left(x'_{E} ,ct'_{E} =0\right)\in {\rm {\mathcal Q}{\mathcal G}}$ is identified with different parameters in the ${\rm {\mathcal S}}$-spacetime, $E=\left(x_{E} ,ct_{E} \right)\in {\rm {\mathcal S}}$.} 
 \label{fig2}
 \end{center}
\end{figure*}
\end{center}

The frames of the causally-biased standard ${\rm {\mathcal S}}$-spacetime and the causally-unbiased ${\rm {\mathcal Q}{\mathcal G}}$-spacetime are depicted in \fref{fig2}. The ${\rm {\mathcal Q}{\mathcal G}}$-spacetime is normalized and illustrated via a noncurved background, with origin $O'=\left(x=0,ct=0\right)$. 

\subsection{Equivalence Transformation for Correlation Measuring }
\begin{theorem}
(Mathematical equivalence of ${\rm {\mathcal Q}{\mathcal G}}$ correlation measuring). Let $f_{g}^{{\rm {\mathcal S}}} \left(A:B\right)$ be an ${\rm {\mathcal S}}$-spacetime correlation measure function, and let $f_{g}^{{\rm {\mathcal Q}{\mathcal G}}} \left(A:B\right)$ be the ${\rm {\mathcal Q}{\mathcal G}}$-spacetime correlation measure between reference event $A:\left\{F_{A} ,Y_{A} \right\}$ and event of interest $B:\left\{F_{B} ,Y_{B}^{i} \right\}$, where $g$ refers to the correlation type. The $\eta $-transformation yields the equivalence $f_{g}^{{\rm {\mathcal Q}{\mathcal G}},\eta } \left(A:B\right)=f_{g}^{{\rm {\mathcal S}}} \left(A:B\right)$, for $\forall g$.
\end{theorem}
\begin{proof}
Let us identify events $A$ and $B$ in the causally-biased space-time ${\rm {\mathcal S}}$, $t_{A} <t_{B} $, $\triangle t=\left|t_{B} -t_{A} \right|>0$ by the measurement/outcome pairs (in particular, event $A$ refers to Alice's correlated local system, while $B$ refers to Bob's correlated local system), as
\begin{equation} \label{ZEqnNum996000} 
A:\left\{F_{A} ,Y_{A} \right\} 
\end{equation} 
and
\begin{equation} \label{ZEqnNum141438} 
B:\left\{F_{B} ,Y_{B}^{i} \right\},                                                   
\end{equation} 
where 
\begin{equation} \label{65)} 
F_{A} =\bigcup _{\forall \varphi _{A} ,y_{A} }\left(x_{A} ,\varphi _{x_{A} } ,y_{x_{A} } \right) ,                                          
\end{equation} 
\begin{equation} \label{66)} 
Y_{A} =\bigcup _{\forall y_{A} }\left(x_{A} ,\varphi _{x_{A} } ,y_{x_{A} } \right) ,                                             
\end{equation} 
while
\begin{equation} \label{67)} 
F_{B} =\bigcup _{\forall \varphi _{B} ,y_{B} }\left(x_{B} ,\varphi _{x_{B} } ,y_{x_{B} } \right) ,                                           
\end{equation} 
\begin{equation} \label{68)} 
Y_{B}^{i} =\bigcup _{\forall y_{B}^{i} }\left(x_{B} ,\varphi _{x_{B} } ,y_{x_{B} }^{i} \right) .                                            
\end{equation} 
The $p_{i}^{{\rm {\mathcal S}}} $ probability of an outcome $Y_{B}^{i} $ of event $B$ with respect to the measurement $F_{B} $ and event $A$ within the ${\rm {\mathcal S}}$-spacetime can be evaluated as
\begin{equation} \label{ZEqnNum176826} 
p_{i}^{{\rm {\mathcal S}}} \left(B\right)=\Pr \left(\left. Y_{B}^{i} \right|Y_{A} ,F_{A} ,F_{B} \right).                                       
\end{equation} 
Specifically, let us evaluate $p_{i}^{{\rm {\mathcal Q}{\mathcal G}}} \left(B\right)$ in the ${\rm {\mathcal Q}{\mathcal G}}$-spacetime. 

Let $t'_{A} =t'_{B} =0$, $\triangle t'=0$, and
\begin{equation} \label{70)} 
A:\left\{F'_{A} ,Y'_{A} \right\},                                                   
\end{equation} 
and
\begin{equation} \label{71)} 
B:\left\{F'_{B} ,{Y'_{B}}^{i} \right\},                                                   
\end{equation} 
where 
\begin{equation} \label{72)} 
F'_{A} =\bigcup _{\forall \varphi _{x'_{A} } ,y_{x'_{A} } }\left(x'_{A} ,\varphi _{x'_{A} } ,y_{x'_{A} } \right) ,                                         
\end{equation} 
\begin{equation} \label{73)} 
Y'_{A} =\bigcup _{\forall y_{x'_{A} } }\left(x'_{A} ,\varphi _{x'_{A} } ,y_{x'_{A} } \right) ,                                           
\end{equation} 
while
\begin{equation} \label{74)} 
F'_{B} =\bigcup _{\forall \varphi _{x'_{B} } ,y_{x'_{B} } }\left(x'_{B} ,\varphi _{x'_{B} } ,y_{x'_{B} } \right) ,                                         
\end{equation} 
\begin{equation} \label{75)} 
{Y'_{B}}^{i}=\bigcup\limits_{\forall {{y}_{x_{B}^{{{i}'}}}}}{\left( {{{{x}'_{B}}}},{{\varphi }_{{{{{x}'_{B}}}}}},{{y}_{x_{B}^{{{i}'}}}} \right)} .                                          
\end{equation} 
Let $x'_{A} $ and $x'_{B} $ be the elementary regions of $A$ and $B$ in the ${\rm {\mathcal Q}{\mathcal G}}$-spacetime, with origin 
\begin{equation} \label{76)} 
O'=\left(x=0,ct\right).                                               
\end{equation} 
At a given $\beta $ strength of the ${\rm {\mathcal Q}{\mathcal G}}$-spacetime relative to the standard space strength $\beta _{{\rm {\mathcal S}}} =0$, $0\le \beta <1$, the elementary region transformation from the ${\rm {\mathcal Q}{\mathcal G}}$-spacetime to the ${\rm {\mathcal S}}$-spacetime is
\begin{equation} \label{77)} 
x_{A} =\eta x'_{A}  
\end{equation} 
and
\begin{equation} \label{78)} 
x_{B} =\eta x'_{B} ,                                                   
\end{equation} 
which yields the $x_{A} $, $x_{B} $ coordinate information of $A$ and $B$ in the ${\rm {\mathcal S}}$-spacetime. Since $x'_{A} $ and $x'_{B} $ are on the $x'$-axis of the ${\rm {\mathcal Q}{\mathcal G}}$-spacetime, parameter $\beta $ can be expressed via the angle $\theta $ of $\left\{x,x'\right\}$, $\left\{ct,ct'\right\}$ of $M$, and $M'$, $0^{\circ } \le \theta <45^{\circ } $, via \eqref{ZEqnNum470950}. 

From the inverse Lorentz transformation, the time difference of events $A$ and $B$ in ${\rm {\mathcal S}}$, $\Delta t=\left|t_{A} -t_{B} \right|$ can be written as
\begin{equation} \label{79)} 
\Delta t=\eta \beta {\textstyle\frac{\Delta x'}{c}} ,                                                
\end{equation} 
where $\Delta x'$ is the location difference of the elementary regions of events $A$ and $B$ in the ${\rm {\mathcal Q}{\mathcal G}}$-spacetime, 
\begin{equation} \label{80)}
\begin{split}
   \Delta {x}'&=\tfrac{1}{\eta }\Delta x \\ 
 & =\left| {{{{x}'_{B}}}}-{{{{x}'_{A}}}} \right| \\ 
 & =\left| \tfrac{1}{\eta }\left( {{x}_{B}}-{{x}_{A}} \right) \right|.  
\end{split}
\end{equation}

Without loss of generality, in the function of $\Delta t$ and $\Delta x'$, the following relation holds for the strength parameter $\beta $:
\begin{equation} \label{81)} 
{\textstyle\frac{\beta ^{2} }{1-\beta ^{2} }} ={\textstyle\frac{\left(\Delta t\right)^{2} }{\left(\Delta x'\right)^{2} }} c^{2} .                                              
\end{equation} 
The transformation is depicted for various strength $\beta $ via $\Delta x'$ in \fref{fig3}. Notation $\Delta x'$ refers to the location difference of $A$ and $B$ in the ${\rm {\mathcal Q}{\mathcal G}}$-spacetime, $O'=\left(x=0,ct\right)$, while $\Delta x$ is the location difference of the same events, $A$ and $B$, in the ${\rm {\mathcal S}}$-spacetime, $\Delta x=\eta \Delta x'$. 

\begin{center}
\begin{figure*}[htbp]
\begin{center}
\includegraphics[angle = 0,width=1\linewidth]{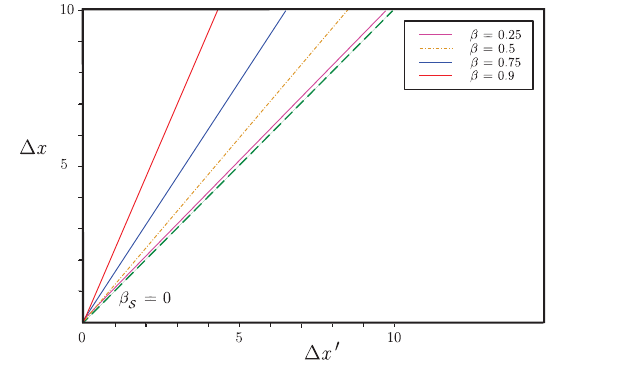}
\caption{The distance $\Delta x$ between $A$ and $B$ in the function of $\Delta x'$, at $\beta =0.25,0.5,0.75,0.9$, $O'=\left(x=0,ct\right)$. At $\beta _{{\rm {\mathcal S}}} =0$, $\Delta x=\Delta x'$.} 
 \label{fig3}
 \end{center}
\end{figure*}
\end{center}

Let $O_{ec} =\left(x_{O_{ec} } =0,ct_{O_{ec} } =0\right)$ refer to an ${\rm {\mathcal S}}$-spacetime event coordinator in the origin of $M$ of the ${\rm {\mathcal S}}$-spacetime, which is causally connected with $A$ via a lightline $L_{1} $ and also causally connected with $B$ via lightline $L_{2} $ in the ${\rm {\mathcal S}}$-spacetime. Practically, $O_{ec} $ sends a flag via $L_{1} $ toward $A$, and a flag via $L_{2} $ to $B$, respectively. When Alice receives the flag, she reveals her system $A$; and when Bob receives it, he reveals his system $B$. 

In the ${\rm {\mathcal Q}{\mathcal G}}$-spacetime, the time information of all events vanishes; therefore, 
\begin{equation} \label{82)} 
t'_{O_{ec} } =t'_{A} =t'_{B} =0.                                              
\end{equation} 
Thus,  
\begin{equation} \label{83)} 
O'_{ec} =\left(x'_{O_{ec} } =0,ct'_{O_{ec} } =0\right)=O'=\left\{x=0,ct\right\},                        
\end{equation} 
and 
\begin{equation} \label{84)} 
A=\left(x'_{A} ,t'_{A} =0\right)\in {\rm {\mathcal Q}{\mathcal G}},                                          
\end{equation} 
\begin{equation} \label{85)} 
B=\left(x'_{B} ,t'_{B} =0\right)\in {\rm {\mathcal Q}{\mathcal G}}.                                          
\end{equation} 
Specifically, since the ${\rm {\mathcal Q}{\mathcal G}}$-spacetime event coordinator $O'_{ec} $ is in the origin of $M'$, $O'_{ec} $ is causally not connected with $A$ and $B$, and there are no existing flags between $O'$, and $A$ and $B$. Consequently, in the ${\rm {\mathcal Q}{\mathcal G}}$-spacetime, $O'_{ec} $ cannot coordinate the events; therefore, $A$ and $B$ are happening simultaneously at $t'_{A} =t'_{B} =0$.

The data transformation between the ${\rm {\mathcal Q}{\mathcal G}}$-spacetime and the ${\rm {\mathcal S}}$-spacetime is summarized in \fref{fig4}. The transformation parameter between frames of ${\rm {\mathcal Q}{\mathcal G}}$ and ${\rm {\mathcal S}}$ is $\eta ={1\mathord{\left/ {\vphantom {1 \sqrt{\left(1-\beta ^{2} \right)} }} \right. \kern-\nulldelimiterspace} \sqrt{\left(1-\beta ^{2} \right)} } $, where $0\le \beta <1$ is the strength coefficient of the ${\rm {\mathcal Q}{\mathcal G}}$-spacetime. The ${\rm {\mathcal Q}{\mathcal G}}$-spacetime has no interpretable background time; $t'_{O_{ec} } =t'_{A} =t'_{B} =0$. The axes are scaled via $\gamma =\sqrt{{\left(1+\beta ^{2} \right)\mathord{\left/ {\vphantom {\left(1+\beta ^{2} \right) \left(1-\beta ^{2} \right)}} \right. \kern-\nulldelimiterspace} \left(1-\beta ^{2} \right)} } $, $l_{scale}^{{\rm {\mathcal Q}{\mathcal G}}} =\gamma l_{scale}^{{\rm {\mathcal S}}} $, $O'_{ec} =\left(x'_{O_{ec} } =0,ct'_{O_{ec} } =0\right)=O'=\left(x=0,ct\right)$.

\begin{center}
\begin{figure*}[htbp]
\begin{center}
\includegraphics[angle = 0,width=1\linewidth]{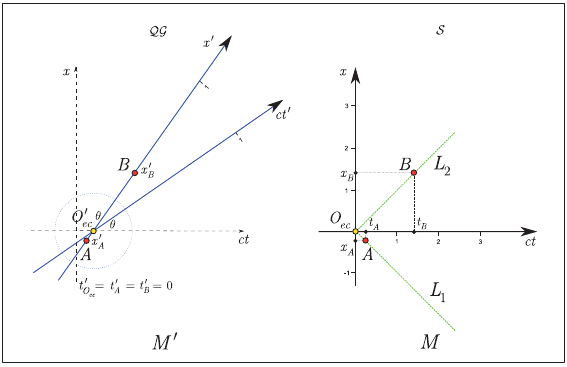}
\caption{Data representation transformation between the causally-unbiased ${\rm {\mathcal Q}{\mathcal G}}$-spacetime ($M'$) with strength coefficient $\beta $, $0\le \beta <1$, and the causally-biased ${\rm {\mathcal S}}$-spacetime ($M$) with reference strength $\beta _{{\rm {\mathcal S}}} =0$. Alice and Bob own their local correlated systems $A$ and $B$. The ${\rm {\mathcal S}}$-spacetime provides a reference frame with axes $x$ and $ct$. In the ${\rm {\mathcal S}}$-spacetime, events $A$ and $B$ are defined by elementary regions $x_{A} $ and $x_{B} $, and the events are happening at time $t_{A} $ and $t_{B} $, $t_{A} <t_{B} $. An event coordinator, $O_{ec} =\left(x_{O_{ec} } =0,ct_{O_{ec} } =0\right)$, sends flags (light beams) to Alice ($A$) and Bob ($B$). When a given party receives the flag, she or he reveals her or his correlated local system. Event $A$ is causally connected with $O_{ec} $ via flag $L_{1} $, and event $B$ is causally connected with $O$ via flag $L_{2} $. In the ${\rm {\mathcal Q}{\mathcal G}}$-spacetime, all time bias vanishes, $t'_{O_{ec} } =t'_{A} =t'_{B} =0$, $O'_{ec} =\left(x'_{O_{ec} } =0,ct'_{O_{ec} } =0\right)=O'=\left(x=0,ct\right)$, $A=\left(x'_{A} ,0\right)$, $B=\left(x'_{A} ,0\right)$. The events in the ${\rm {\mathcal Q}{\mathcal G}}$-spacetime are causally not connected with the event coordinator $O'_{ec} $; thus, there is no flag (light beam) that connects them. At a given $\beta $, $\beta =\tan \left(\theta \right)$, $0^{\circ } \le \theta <45^{\circ } $, the ${\rm {\mathcal Q}{\mathcal G}}$-spacetime elementary regions of $A$ and $B$ are ${{{x}'_{A}}}={{{x}_{A}}}/{\eta }\;$, ${{{x}'_{B}}}={{{x}_{B}}}/{\eta }\;$, where $\eta ={1}/{\sqrt{\left( 1-{{\beta }^{2}} \right)}}\;$. The axes are scaled via $l_{scale}^{\mathcal{Q}\mathcal{G}}=\gamma l_{scale}^{\mathcal{S}}$, $\gamma =\sqrt{{\left( 1+{{\beta }^{2}} \right)}/{\left( 1-{{\beta }^{2}} \right)}\;}$.} 
 \label{fig4}
 \end{center}
\end{figure*}
\end{center}

The entropy function for the $\eta $-transformed ${\rm {\mathcal Q}{\mathcal G}}$-spacetime is derived as follows. Let $p_{i}^{{\rm {\mathcal Q}{\mathcal G}},\eta } $ refer to the probability with the $\eta $-transformed $\eta x'_{A} =x_{A} $, $\eta x'_{B} =x_{B} $ elementary regions. Then, the $p_{i}^{{\rm {\mathcal Q}{\mathcal G}},\eta } $ probability of an outcome ${Y'_{B}}^{i}$ of event $B$ with respect to measurement $F'_{B} $ and event $A$ is evaluated as
\begin{equation} \label{ZEqnNum860456} 
p_{i}^{{\rm {\mathcal Q}{\mathcal G}},\eta } \left(B\right)=\Pr \left(\left. Y_{B,\eta }^{i'} \right|Y'_{A,\eta } ,F'_{A,\eta } ,F'_{B,\eta } \right),                                     
\end{equation} 
where
\begin{equation} \label{87)}
\begin{split}
   {{{{F}'_{A,\eta }}}}&=\bigcup\limits_{\forall {{\varphi }_{\eta {{{{x}'_{A}}}}}},{{y}_{\eta {{{{x}'_{A}}}}}}}{\left( \eta {{{{x}'_{A}}}},{{\varphi }_{\eta {{{{x}'_{A}}}}}},{{y}_{\eta {{{{x}'_{A}}}}}} \right)} \\ 
 & =\bigcup\limits_{\forall {{\varphi }_{{{x}_{A}}}},{{y}_{{{x}_{A}}}}}{\left( {{x}_{A}},{{\varphi }_{{{x}_{A}}}},{{y}_{{{x}_{A}}}} \right)},  
\end{split}
\end{equation}
\begin{equation} \label{88)}
\begin{split}
   {{{{Y}'_{A,\eta }}}}&=\bigcup\limits_{\forall {{y}_{\eta {{{{x}'_{A}}}}}}}{\left( \eta {{{{x}'_{A}}}},{{\varphi }_{\eta {{{{x}'_{A}}}}}},{{y}_{\eta {{{{x}'_{A}}}}}} \right)} \\ 
 & =\bigcup\limits_{\forall {{y}_{{{x}_{A}}}}}{\left( {{x}_{A}},{{\varphi }_{{{x}_{A}}}},{{y}_{{{x}_{A}}}} \right)},  
\end{split}
\end{equation}
and
\begin{equation} \label{89)}
\begin{split}
   {{{{F}'_{B,\eta }}}}&=\bigcup\limits_{\forall {{\varphi }_{\eta {{{{x}'_{B}}}}}},{{y}_{\eta {{{{x}'_{B}}}}}}}{\left( \eta {{{{x}'_{B}}}},{{\varphi }_{\eta {{{{x}'_{B}}}}}},{{y}_{\eta {{{{x}'_{B}}}}}} \right)} \\ 
 & =\bigcup\limits_{\forall {{\varphi }_{{{x}_{B}}}},{{y}_{{{x}_{B}}}}}{\left( {{x}_{B}},{{\varphi }_{{{x}_{B}}}},{{y}_{{{x}_{B}}}} \right)},  
\end{split}
\end{equation}
\begin{equation} \label{90)}
\begin{split}
   Y_{B,\eta }^{{{i}'}}&=\bigcup\limits_{\forall {{y}_{\eta {{{{x}'_{B}}}}}}}{\left( \eta {{{{x}'_{B}}}},{{\varphi }_{\eta {{{{x}'_{B}}}}}},{{y}_{\eta {{{{x}'_{B}}}}}} \right)} \\ 
 & =\bigcup\limits_{\forall {{y}_{{{x}_{B}}}}}{\left( {{x}_{B}},{{\varphi }_{{{x}_{B}}}},{{y}_{{{x}_{B}}}} \right)}.  
\end{split}
\end{equation}

Particularly, $p_{i}^{{\rm {\mathcal Q}{\mathcal G}},\eta } \left(B\right)$ can be expressed via $p_{i}^{{\rm {\mathcal S}}} \left(B\right)$ of ${\rm {\mathcal S}}$ using the $\eta $-transformed $x'_{A} $ and $x'_{B} $ of ${\rm {\mathcal Q}{\mathcal G}}$. This connection allows us to construct the $f_{g}^{{\rm {\mathcal Q}{\mathcal G}},\eta } \left(\cdot \right)$ correlation measure functions of the ${\rm {\mathcal Q}{\mathcal G}}$-spacetime with $\eta $-transformed arguments, leading to identical mathematical structures as the $f_{g}^{{\rm {\mathcal S}}} \left(A:B\right)$ ${\rm {\mathcal S}}$-spacetime correlation measure functions; thus without loss of generality,
\begin{equation} \label{ZEqnNum789082} 
p_{i}^{{\rm {\mathcal Q}{\mathcal G}},\eta } \left(\cdot \right)=p_{i}^{{\rm {\mathcal S}}} \left(\cdot \right).                                              
\end{equation} 
Too see it, let $f_{g} \left(\cdot \right)$ identify the entropy function, $f_{g} \left(\cdot \right)=H\left(\cdot \right)$. Then let $H^{{\rm {\mathcal S}}} \left(\cdot \right)$, $H^{{\rm {\mathcal Q}{\mathcal G}},\eta } \left(\cdot \right)$ refer to the ${\rm {\mathcal S}}$-spacetime function and the ${\rm {\mathcal Q}{\mathcal G}}$-spacetime entropy function with $\eta $-transformed arguments, respectively. 

Let $H^{{\rm {\mathcal S}}} \left(B\right)$ be the entropy of $B$ relative to $A$, $t_{A} <t_{B} $ in ${\rm {\mathcal S}}$, as 
\begin{equation} \label{92)} 
H^{{\rm {\mathcal S}}} \left(B\right)=-\sum _{i}p_{i}^{{\rm {\mathcal S}}}  \log _{2} p_{i}^{{\rm {\mathcal S}}} ,                                       
\end{equation} 
where $p_{i}^{{\rm {\mathcal S}}} $ is given in \eqref{ZEqnNum176826}; thus,
\begin{equation} \label{93)} 
H^{{\rm {\mathcal S}}} \left(B\right)=-\sum _{i}\Pr \left(\left. Y_{B}^{i} \right|Y_{A} ,F_{A} ,F_{B} \right) \log _{2} \Pr \left(\left. Y_{B}^{i} \right|Y_{A} ,F_{A} ,F_{B} \right) 
\end{equation} 
and
\begin{equation} \label{94)} 
H^{{\rm {\mathcal Q}{\mathcal G}},\eta } \left(B\right)=-\sum _{i}p_{i}^{{\rm {\mathcal Q}{\mathcal G}},\eta }  \log _{2} p_{i}^{{\rm {\mathcal Q}{\mathcal G}},\eta } ,                                   
\end{equation} 
while $p_{i}^{{\rm {\mathcal Q}{\mathcal G}},\eta } $ is given in \eqref{ZEqnNum860456}. 

Without loss of generality, $H^{{\rm {\mathcal Q}{\mathcal G}},\eta } \left(B\right)$ is evaluated as
\begin{equation}\label{95)}
\begin{split}
  & {{H}^{\mathcal{Q}\mathcal{G},\eta }}\left( B \right)\\
  &=-\sum\limits_{i}{\Pr \left( \left. Y_{B,\eta }^{{i}'} \right|{{{{Y}'_{A,\eta }}}},{{{{F}'_{A,\eta }}}},{{{{F}'_{B,\eta }}}} \right)}{{\log }_{2}}\Pr \left( \left. Y_{B,\eta }^{{i}'} \right|{{{{Y}'_{A,\eta }}}},{{{{F}'_{A,\eta }}}},{{{{F}'_{B,\eta }}}} \right) \\ 
 & =-\sum\limits_{i}{\Pr \left( \begin{split}
  & \left. \bigcup\limits_{\forall {{y}_{\eta {{{{x}'_{B}}}}}}}{\left( \eta {{{{x}'_{B}}}},{{\varphi }_{\eta {{{{x}'_{B}}}}}},{{y}_{\eta {{{{x}'_{B}}}}}} \right)} \right|\bigcup\limits_{\forall {{y}_{\eta {{{{x}'_{A}}}}}}}{\left( \eta {{{{x}'_{A}}}},{{\varphi }_{\eta {{{{x}'_{A}}}}}},{{y}_{\eta {{{{x}'_{A}}}}}} \right)}, \\ 
 & \bigcup\limits_{\forall {{\varphi }_{\eta {{{{x}'_{A}}}}}},{{y}_{\eta {{{{x}'_{A}}}}}}}{\left( \eta {{{{x}'_{A}}}},{{\varphi }_{\eta {{{{x}'_{A}}}}}},{{y}_{\eta {{{{x}'_{A}}}}}} \right)},\bigcup\limits_{\forall \varphi \eta {{{{x}'_{B}}}},{{y}_{\eta {{{{x}'_{B}}}}}}}{\left( \eta {{{{x}'_{B}}}},{{\varphi }_{\eta {{{{x}'_{B}}}}}},{{y}_{\eta {{{{x}'_{B}}}}}} \right)} \\ 
\end{split} \right)} \\ 
& \text{  }\cdot {{\log }_{2}}\Pr \left( \begin{split}
  & \left. \bigcup\limits_{\forall {{y}_{\eta {{{{x}'_{B}}}}}}}{\left( \eta {{{{x}'_{B}}}},{{\varphi }_{\eta {{{{x}'_{B}}}}}},{{y}_{\eta {{{{x}'_{B}}}}}} \right)} \right|\bigcup\limits_{\forall {{y}_{\eta {{{{x}'_{A}}}}}}}{\left( \eta {{{{x}'_{A}}}},{{\varphi }_{\eta {{{{x}'_{A}}}}}},{{y}_{\eta {{{{x}'_{A}}}}}} \right)}, \\ 
 & \bigcup\limits_{\forall {{\varphi }_{\eta {{{{x}'_{A}}}}}},{{y}_{\eta {{{{x}'_{A}}}}}}}{\left( \eta {{{{x}'_{A}}}},{{\varphi }_{\eta {{{{x}'_{A}}}}}},{{y}_{\eta {{{{x}'_{A}}}}}} \right)},\bigcup\limits_{\forall {{\varphi }_{\eta {{{{x}'_{B}}}}}},{{y}_{\eta {{{{x}'_{B}}}}}}}{\left( \eta {{{{x}'_{B}}}},{{\varphi }_{\eta {{{{x}'_{B}}}}}},{{y}_{\eta {{{{x}'_{B}}}}}} \right)} \\ 
\end{split} \right) \\ 
 & =-\sum\limits_{i}{\Pr \left( \begin{split}
  & \left. \bigcup\limits_{\forall {{y}_{B}}}{\left( {{x}_{B}},{{\varphi }_{{{x}_{B}}}},{{y}_{{{x}_{B}}}} \right)} \right|\bigcup\limits_{\forall {{y}_{A}}}{\left( {{x}_{A}},{{\varphi }_{{{x}_{A}}}},{{y}_{{{x}_{A}}}} \right)}, \\ 
 & \bigcup\limits_{\forall {{\varphi }_{A}},{{y}_{A}}}{\left( {{x}_{A}},{{\varphi }_{{{x}_{A}}}},{{y}_{{{x}_{A}}}} \right)},\bigcup\limits_{\forall {{\varphi }_{B}},{{y}_{B}}}{\left( {{x}_{B}},{{\varphi }_{{{x}_{B}}}},{{y}_{{{x}_{B}}}} \right)} \\ 
\end{split} \right)} \\ 
 & \text{  }\cdot {{\log }_{2}}\Pr \left( \begin{split}
  & \left. \bigcup\limits_{\forall {{y}_{B}}}{\left( {{x}_{B}},{{\varphi }_{{{x}_{B}}}},{{y}_{{{x}_{B}}}} \right)} \right|\bigcup\limits_{\forall {{y}_{A}}}{\left( {{x}_{A}},{{\varphi }_{{{x}_{A}}}},{{y}_{{{x}_{A}}}} \right)}, \\ 
 & \bigcup\limits_{\forall {{\varphi }_{A}},{{y}_{A}}}{\left( {{x}_{A}},{{\varphi }_{{{x}_{A}}}},{{y}_{{{x}_{A}}}} \right)},\bigcup\limits_{\forall {{\varphi }_{B}},{{y}_{B}}}{\left( {{x}_{B}},{{\varphi }_{{{x}_{B}}}},{{y}_{{{x}_{B}}}} \right)} \\ 
\end{split} \right) \\ 
 & =-\sum\limits_{i}{\Pr \left( \left. Y_{B}^{i} \right|{{Y}_{A}},{{F}_{A}},{{F}_{B}} \right)}{{\log }_{2}}\Pr \left( \left. Y_{B}^{i} \right|{{Y}_{A}},{{F}_{A}},{{F}_{B}} \right) \\ 
 & ={{H}^{\mathcal{S}}}\left( B \right).  
\end{split}
\end{equation} 

In particular, function $H^{{\rm {\mathcal Q}{\mathcal G}},\eta } \left(B\right)$, with the $\eta $-transformed $x'_{A} $ and $x'_{B} $ arguments of the ${\rm {\mathcal Q}{\mathcal G}}$-spacetime is defined by same mathematical background as $H^{{\rm {\mathcal S}}} \left(B\right)$ with inputs $x_{A} $ and $x_{B} $ from the ${\rm {\mathcal S}}$-spacetime. As a corollary, the entropy calculations in the ${\rm {\mathcal Q}{\mathcal G}}$-spacetime can be performed via the mathematical properties of $H^{{\rm {\mathcal S}}} \left(\cdot \right)$ of the ${\rm {\mathcal S}}$-spacetime. 

Since for $\forall g$, $p_{i}^{{\rm {\mathcal S}}} $ is a fundament of the $f_{g}^{{\rm {\mathcal S}}} \left(\cdot \right)$ correlation measure of the ${\rm {\mathcal S}}$-spacetime, it follows that
\begin{equation}\label{ZEqnNum968507}
f_{g}^{{\rm {\mathcal Q}{\mathcal G}},\eta } \left(\cdot \right)=f_{g}^{{\rm {\mathcal S}}} \left(\cdot \right),
\end{equation}
for $\forall g$.

The proof is concluded here.
\end{proof}

\subsection{Extension of Composite Regions}
\begin{lemma}
The $\eta $-transformation extends the sets of elementary regions. 
\end{lemma}
\begin{proof}
Let ${\rm {\mathcal O}}_{C'} \in {\rm {\mathcal Q}{\mathcal G}}$ be a composite region in ${\rm {\mathcal Q}{\mathcal G}}$, and let $R_{C'} $ be the union of all sets of $R_{x'} $ measurement information for the $x'\in {\rm {\mathcal O}}_{C'} $ elementary regions and
\begin{equation} \label{97)} 
R_{C'} =\bigcup _{x'\in {\rm {\mathcal O}}_{C'} }R_{x'}  .                                              
\end{equation} 
Let ${\rm {\mathcal O}}_{C} \in {\rm {\mathcal S}}$ be a composite region in ${\rm {\mathcal S}}$, while $R_{x} $ refers to the measurement information for the elementary region $x\in {\rm {\mathcal O}}_{C} $; then 
\begin{equation} \label{98)} 
R_{C} =\bigcup _{x\in {\rm {\mathcal O}}_{C} }R_{x}  .                                               
\end{equation} 
Then let $V'$ for the ${\rm {\mathcal Q}{\mathcal G}}$-spacetime be defined as
\begin{equation} \label{99)} 
V'=\bigcup _{\forall x'}R_{x'}  ,                                                
\end{equation} 
and let $V$ for the ${\rm {\mathcal S}}$-spacetime be defined as
\begin{equation} \label{100)} 
V=\bigcup _{\forall x}R_{x}  .                                                 
\end{equation} 
For any nonzero $\eta $, $\Delta x=\eta \Delta x'$, the composite region ${\rm {\mathcal O}}_{C'} $ is extended onto ${\rm {\mathcal O}}_{C} $; thus,
\begin{equation} \label{101)} 
R_{C'} \ne R_{C} .                                                 
\end{equation} 
Let $F'_{x} $ refer to a procedure in a region $x'\in {\rm {\mathcal O}}_{C'} $, 
\begin{equation} \label{102)} 
F_{C'} =\bigcup _{x'\in {\rm {\mathcal O}}_{C'} }F_{x'}  .                                             
\end{equation} 
For any nonzero $\eta $, the functions are reevaluated for ${\rm {\mathcal O}}_{C} $ as
\begin{equation} \label{103)} 
F_{C} =\bigcup _{x'\eta \in {\rm {\mathcal O}}_{C} }F_{x'\eta }  .                                           
\end{equation} 
Let $Y'_{x} $ refer to the outcome set for a region $x'\in {\rm {\mathcal O}}_{C'} $, 
\begin{equation} \label{104)} 
Y_{C'} =\bigcup _{x'\in {\rm {\mathcal O}}_{C'} }Y_{x'}  ;                                            
\end{equation} 
then, for any nonzero $\eta $, the $Y_{C} $ set for $x\in {\rm {\mathcal O}}_{C} $ is
\begin{equation} \label{105)} 
Y_{C} =\bigcup _{\eta x'\in {\rm {\mathcal O}}_{C} }Y_{x}  .                                             
\end{equation} 
Recall that for the composite region ${\rm {\mathcal O}}_{C'} $,
\begin{equation} \label{106)} 
Y_{C'} \subseteq F_{C'} \subseteq R_{C'} \in {\rm {\mathcal Q}{\mathcal G}},                                      
\end{equation} 
and for the extended composite region ${\rm {\mathcal O}}_{C} $, 
\begin{equation} \label{107)} 
Y_{C} \subseteq F_{C} \subseteq R_{C} \in {\rm {\mathcal S}}.                                      
\end{equation} 
Specifically, putting the pieces together, the corresponding relations of the sets are as
\begin{equation} \label{108)} 
R_{C} \supseteq R_{C'} , 
\end{equation} 
and
\begin{equation} \label{109)} 
\begin{array}{l} {F_{C} \supseteq F_{C'} ,} \\ {Y_{C} \supseteq Y_{C'} .} \end{array} 
\end{equation} 
Particularly, for any nonzero $\eta $, the relation of the sets of the ${\rm {\mathcal S}}$-spacetime and ${\rm {\mathcal Q}{\mathcal G}}$-spacetime is as
\begin{equation} \label{ZEqnNum139890} 
Y_{C'} \subseteq Y_{C} \subseteq F_{C'} \subseteq F_{C} \subseteq R_{C'} \subseteq R_{C} .                                   
\end{equation} 
\end{proof}

\begin{center}
\begin{figure*}[htbp]
\begin{center}
\includegraphics[angle = 0,width=1\linewidth]{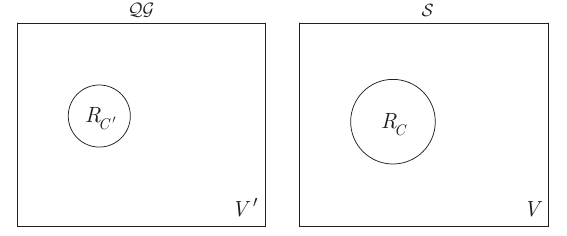}
\caption{The result of the elementary region transformation. The region $R_{C'} $ of the ${\rm {\mathcal Q}{\mathcal G}}$-spacetime is transformed into the region $R_{C} $ in the ${\rm {\mathcal Q}{\mathcal G}}$-spacetime via a nonzero $\eta $.} 
 \label{fig5}
 \end{center}
\end{figure*}
\end{center}

The $R_{C'} \in {\rm {\mathcal Q}{\mathcal G}}$ with ${\rm {\mathcal O}}_{C'} $, at any nonzero $\eta $, leads to $R_{C} \in {\rm {\mathcal S}}$, with the exposed composite region ${\rm {\mathcal O}}_{C} $, as illustrated in \fref{fig5}. 

\section{Correlation Measure Functions}
\label{sec4}
\begin{theorem}
The $f_{g}^{{\rm {\mathcal Q}{\mathcal G}},\eta } \left(\cdot \right)$ functions determine the correlations in the ${\rm {\mathcal Q}{\mathcal G}}$-spacetime through the causally-biased functions of the ${\rm {\mathcal S}}$-spacetime for $\forall g$. 
\end{theorem}
\begin{proof}
Let $A$ and $B$, $A:\left\{F'_{A} ,Y'_{A} \right\}\in {\rm {\mathcal Q}{\mathcal G}}$, $B:\left\{F'_{B} ,Y_{B}^{i'} \right\}\in {\rm {\mathcal Q}{\mathcal G}}$, with ${\rm {\mathcal S}}$-spacetime relation $t_{A} <t_{B} $, and ${\rm {\mathcal Q}{\mathcal G}}$-spacetime relation $t'_{A} =t'_{B} =0$, refer to the input and output of an ${\rm {\mathcal N}}_{AB} $ noisy process in the ${\rm {\mathcal Q}{\mathcal G}}$-spacetime, respectively. 

Based on Theorem 1, the correlation measures are evaluated as follows.  

Let corresponding probabilities of the $\eta $-transformed ${\rm {\mathcal Q}{\mathcal G}}$-spacetime be referred as
\begin{equation} \label{ZEqnNum990411} 
p_{j}^{{\rm {\mathcal Q}{\mathcal G}},\eta } \left(A\right)=\Pr \left(\left. Y_{A,\eta }^{j'} \right|Y'_{B,\eta } ,F'_{B,\eta } ,F'_{A,\eta } \right),                                  
\end{equation} 
\begin{equation} \label{ZEqnNum313223} 
p_{i}^{{\rm {\mathcal Q}{\mathcal G}},\eta } \left(B\right)=\Pr \left(\left. Y_{B,\eta }^{i'} \right|Y'_{A,\eta } ,F'_{A,\eta } ,F'_{B,\eta } \right),                                  
\end{equation} 
and
\begin{equation} \label{ZEqnNum881654} 
p_{i,j}^{{\rm {\mathcal Q}{\mathcal G}},\eta } \left(A,B\right)=\Pr \left(\left. Y_{A,\eta }^{i'} ,Y_{B,\eta }^{j'} \right|Y'_{A,\eta } ,Y'_{B,\eta } ,F'_{A,\eta } ,F'_{B,\eta } \right).                        
\end{equation} 
In particular, using \eqref{ZEqnNum990411}, \eqref{ZEqnNum313223}, and \eqref{ZEqnNum881654}, the $I^{{\rm {\mathcal Q}{\mathcal G}},\eta } \left(A:B\right)$ mutual information between the $\eta $-transformed arguments of $A$ and $B$ is expressed via entropy $H^{{\rm {\mathcal Q}{\mathcal G}},\eta } \left(B\right)$, and the conditional entropy $H^{{\rm {\mathcal Q}{\mathcal G}},\eta } \left(\left. B\right|A\right)$ as
\begin{equation}\label{114)}
\begin{split}
   {{I}^{\mathcal{Q}\mathcal{G},\eta }}\left( A:B \right)&={{H}^{\mathcal{Q}\mathcal{G},\eta }}\left( B \right)-{{H}^{\mathcal{Q}\mathcal{G},\eta }}\left( \left. B \right|A \right) \\ 
 & =-\sum\limits_{i}{p_{i}^{\mathcal{Q}\mathcal{G},\eta }}\left( B \right){{\log }_{2}}p_{i}^{\mathcal{Q}\mathcal{G},\eta }\left( B \right)-\left( \sum\limits_{i,j}{p_{i,j}^{\mathcal{Q}\mathcal{G},\eta }}\left( A,B \right){{\log }_{2}}\tfrac{p_{i}^{\mathcal{Q}\mathcal{G},\eta }\left( A \right)}{p_{i,j}^{\mathcal{Q}\mathcal{G},\eta }\left( A,B \right)} \right) \\ 
 & =-\sum\limits_{i}{\Pr \left( \left. Y_{B,\eta }^{{i}'} \right|{{{{Y}'_{A,\eta }}}},{{{{F}'_{A,\eta }}}},{{{{F}'_{B,\eta }}}} \right)}{{\log }_{2}}\Pr \left( \left. Y_{A,\eta }^{{j}'} \right|{{{{Y}'_{B,\eta }}}},{{{{F}'_{B,\eta }}}},{{{{F}'_{A,\eta }}}} \right) \\ 
 & \text{  }-\left( \sum\limits_{i,j}{\Pr \left( \left. Y_{A,\eta }^{{i}'},Y_{B,\eta }^{{j}'} \right|{{{{Y}'_{A,\eta }}}},{{{{Y}'_{B,\eta }}}},{{{{F}'_{A,\eta }}}},{{{{F}'_{B,\eta }}}} \right)}{{\log }_{2}}\tfrac{\Pr \left( \left. Y_{A,\eta }^{{j}'} \right|{{{{Y}'_{B,\eta }}}},{{{{F}'_{B,\eta }}}},{{{{F}'_{A,\eta }}}} \right)}{\Pr \left( \left. Y_{A,\eta }^{{i}'},Y_{B,\eta }^{{j}'} \right|{{{{Y}'_{A,\eta }}}},{{{{Y}'_{B,\eta }}}},{{{{F}'_{A,\eta }}}},{{{{F}'_{B,\eta }}}} \right)} \right).  
\end{split}
\end{equation} 

The $\chi ^{{\rm {\mathcal Q}{\mathcal G}},\eta } \left(A:B\right)$ Holevo quantity with the $\eta $-transformed arguments of $A$ and $B$ is evaluated via \eqref{ZEqnNum990411}, \eqref{ZEqnNum313223}, and \eqref{ZEqnNum881654} as
\begin{equation} \label{115)} 
\chi ^{{\rm {\mathcal Q}{\mathcal G}},\eta } \left(A:B\right)={\rm S}\left({\rm {\mathcal N}}_{AB} \left(\sum _{i}p_{i}^{{\rm {\mathcal Q}{\mathcal G}},\eta } \rho _{i}^{{\rm {\mathcal Q}{\mathcal G}},\eta }  \right)\right)-\sum _{i}p_{i}^{{\rm {\mathcal Q}{\mathcal G}},\eta } {\rm S}\left({\rm {\mathcal N}}_{AB} \left(\rho _{i}^{{\rm {\mathcal Q}{\mathcal G}},\eta } \right)\right) , 
\end{equation} 
where $\rho _{i}^{{\rm {\mathcal Q}{\mathcal G}},\eta } $ is as
\begin{equation} \label{116)} 
\rho _{i}^{{\rm {\mathcal Q}{\mathcal G}},\eta } =\sum _{k=0}^{n-1}p_{k}^{{\rm {\mathcal Q}{\mathcal G}},\eta }  {\left| x_{k}  \right\rangle} {\left\langle x_{k}  \right|} , 
\end{equation} 
and ${\rm S}\left(\cdot \right)$ is the quantum entropy expressed for a density matrix $\rho $ as
\begin{equation} \label{117)} 
{\rm S}\left(\rho \right)=-\text{Tr}\left(\rho \log \left(\rho \right)\right).                                        
\end{equation} 
The $I_{coh}^{{\rm {\mathcal Q}{\mathcal G}},\eta } \left(A:B\right)$ coherent information is expressed via $H^{{\rm {\mathcal Q}{\mathcal G}},\eta } \left(\cdot \right)$ as
\begin{equation}\label{118)}
\begin{split}
   I_{coh}^{\mathcal{Q}\mathcal{G},\eta }\left( A:B \right)&=-{{H}^{\mathcal{Q}\mathcal{G},\eta }}\left( \left. A \right|B \right) \\ 
 & =-\left( \sum\limits_{i,j}{p_{i,j}^{\mathcal{Q}\mathcal{G},\eta }}\left( A,B \right){{\log }_{2}}\tfrac{p_{j}^{\mathcal{Q}\mathcal{G},\eta }\left( B \right)}{p_{i,j}^{\mathcal{Q}\mathcal{G},\eta }\left( A,B \right)} \right) \\ 
 & =-\left( \sum\limits_{i,j}{\Pr \left( \left. Y_{A,\eta }^{{i}'},Y_{B,\eta }^{{j}'} \right|{{{{Y}'_{A,\eta }}}},{{{{Y}'_{B,\eta }}}},{{{{F}'_{A,\eta }}}},{{{{F}'_{B,\eta }}}} \right)}{{\log }_{2}}\tfrac{\Pr \left( \left. Y_{B,\eta }^{{i}'} \right|{{{{Y}'_{A,\eta }}}},{{{{F}'_{A,\eta }}}},{{{{F}'_{B,\eta }}}} \right)}{\Pr \left( \left. Y_{A,\eta }^{{i}'},Y_{B,\eta }^{{j}'} \right|{{{{Y}'_{A,\eta }}}},{{{{Y}'_{B,\eta }}}},{{{{F}'_{A,\eta }}}},{{{{F}'_{B,\eta }}}} \right)} \right).  
\end{split}
\end{equation} 

Particularly, from the derivation of the correlation measure quantities $I^{{\rm {\mathcal Q}{\mathcal G}},\eta } \left(\cdot \right)$, $\chi ^{{\rm {\mathcal Q}{\mathcal G}},\eta } \left(\cdot \right)$, $I_{coh}^{{\rm {\mathcal Q}{\mathcal G}},\eta } \left(\cdot \right)$ of the ${\rm {\mathcal Q}{\mathcal G}}$-spacetime, it follows that 
\begin{equation} \label{ZEqnNum300250} 
I^{{\rm {\mathcal Q}{\mathcal G}},\eta } \left(A:B\right)=I^{{\rm {\mathcal S}}} \left(A:B\right),                                      
\end{equation} 
\begin{equation} \label{ZEqnNum936477} 
\chi ^{{\rm {\mathcal Q}{\mathcal G}},\eta } \left(A:B\right)=\chi ^{{\rm {\mathcal S}}} \left(A:B\right),                                      
\end{equation} 
and
\begin{equation} \label{ZEqnNum891432} 
I_{coh}^{{\rm {\mathcal Q}{\mathcal G}},\eta } \left(A:B\right)=I_{coh}^{{\rm {\mathcal S}}} \left(A:B\right),                                     
\end{equation} 
where $I^{{\rm {\mathcal S}}} \left(\cdot \right)$, $\chi ^{{\rm {\mathcal S}}} \left(\cdot \right)$, $I_{coh}^{{\rm {\mathcal S}}} \left(\cdot \right)$ stand for the ${\rm {\mathcal S}}$-spacetime mutual information, Holevo information and coherent information, respectively.

Let $C^{{\rm {\mathcal Q}{\mathcal G}},\eta } \left({\rm {\mathcal N}}_{AB} \right)$ refer to the classical capacity of ${\rm {\mathcal N}}_{AB} $ with the $\eta $-transformed arguments of the ${\rm {\mathcal Q}{\mathcal G}}$-spacetime, the $C_{{\rm {\mathcal X}}}^{{\rm {\mathcal Q}{\mathcal G}},\eta } \left({\rm {\mathcal N}}_{AB} \right)$ Holevo capacity of ${\rm {\mathcal N}}_{AB} $ with the $\eta $-transformed arguments of the ${\rm {\mathcal Q}{\mathcal G}}$-spacetime, the $P^{{\rm {\mathcal Q}{\mathcal G}},\eta } \left({\rm {\mathcal N}}_{AB} \right)$ private classical capacity of ${\rm {\mathcal N}}_{AB} $ with the $\eta $-transformed arguments of the ${\rm {\mathcal Q}{\mathcal G}}$-spacetime, and the $Q^{{\rm {\mathcal Q}{\mathcal G}},\eta } \left({\rm {\mathcal N}}_{AB} \right)$ quantum capacity of ${\rm {\mathcal N}}_{AB} $ with the $\eta $-transformed arguments of the ${\rm {\mathcal Q}{\mathcal G}}$-spacetime, respectively. 

Without loss of generality, from \eqref{ZEqnNum300250}, \eqref{ZEqnNum936477}, and \eqref{ZEqnNum891432}, it follows that the capacity measures of ${\rm {\mathcal N}}_{AB} $ in the ${\rm {\mathcal Q}{\mathcal G}}$-spacetime are evaluated via the ${\rm {\mathcal S}}$-spacetime functions as 
\begin{equation} \label{122)} 
C^{{\rm {\mathcal Q}{\mathcal G}},\eta } \left({\rm {\mathcal N}}_{AB} \right)=C^{{\rm {\mathcal S}}} \left({\rm {\mathcal N}}_{AB} \right), 
\end{equation} 
\begin{equation} \label{123)} 
C_{{\rm {\mathcal X}}}^{{\rm {\mathcal Q}{\mathcal G}},\eta } \left({\rm {\mathcal N}}_{AB} \right)=C_{{\rm {\mathcal X}}}^{{\rm {\mathcal S}}} \left({\rm {\mathcal N}}_{AB} \right), 
\end{equation} 
\begin{equation} \label{124)} 
P^{{\rm {\mathcal Q}{\mathcal G}},\eta } \left({\rm {\mathcal N}}_{AB} \right)=P^{{\rm {\mathcal S}}} \left({\rm {\mathcal N}}_{AB} \right), 
\end{equation} 
\begin{equation} \label{125)} 
Q^{{\rm {\mathcal Q}{\mathcal G}},\eta } \left({\rm {\mathcal N}}_{AB} \right)=Q^{{\rm {\mathcal S}}} \left({\rm {\mathcal N}}_{AB} \right), 
\end{equation} 
where $C^{{\rm {\mathcal S}}} \left({\rm {\mathcal N}}_{AB} \right),$ $C_{{\rm {\mathcal X}}}^{{\rm {\mathcal S}}} \left({\rm {\mathcal N}}_{AB} \right),$ $P^{{\rm {\mathcal S}}} \left({\rm {\mathcal N}}_{AB} \right),$ $Q^{{\rm {\mathcal S}}} \left({\rm {\mathcal N}}_{AB} \right)$ refer to the capacities of ${\rm {\mathcal N}}_{AB} $ in the ${\rm {\mathcal S}}$-spacetime. 

Specifically, the results can be extended to arbitrary correlation measure functions via $p^{{\rm {\mathcal Q}{\mathcal G}},\eta } \left(\cdot \right)$, the relation $f_{g}^{{\rm {\mathcal Q}{\mathcal G}},\eta } \left(\cdot \right)=f_{g}^{{\rm {\mathcal S}}} \left(\cdot \right)$, for $\forall g$, is straightforwardly follows from \eqref{ZEqnNum968507}.
\end{proof}
 \section{Conclusions}
\label{sec5}
In the quantum gravity space, the causal structure is dynamic and all time information vanishes, which makes it no possible to utilize the causally-biased correlation measure functions. Since the correction of the causally-unbiased entropy function has led to an obscure mathematical structure, we introduced a data representation correction, which makes it possible to apply the causally-biased functions in a causally-unbiased structure. The equivalence transformation uses a correction in the data representation of the causally-unbiased quantum gravity space, which has a consequence that all mathematical properties of the causally-biased entropy function are preserved in the causally-unbiased space. The proposed transformation allows us to calculate correlations in the quantum gravity space with the stable mathematical background and apparatus of the causally-biased correlation measure functions. We demonstrated the results through the causally-unbiased probability, entropy and correlation measure functions.

\section*{Acknowledgements}
The research reported in this paper has been supported by the Hungarian Academy of Sciences (MTA Premium Postdoctoral Research Program 2019), by the National Research, Development and Innovation Fund (TUDFO/51757/2019-ITM, Thematic Excellence Program), by the National Research Development and Innovation Office of Hungary (Project No. 2017-1.2.1-NKP-2017-00001), by the Hungarian Scientific Research Fund - OTKA K-112125 and in part by the BME Artificial Intelligence FIKP grant of EMMI (Budapest University of Technology, BME FIKP-MI/SC).


\newpage
\appendix
\setcounter{table}{0}
\setcounter{figure}{0}
\setcounter{equation}{0}
\setcounter{algocf}{0}
\renewcommand{\thetable}{\Alph{section}.\arabic{table}}
\renewcommand{\thefigure}{\Alph{section}.\arabic{figure}}
\renewcommand{\theequation}{\Alph{section}.\arabic{equation}}
\renewcommand{\thealgocf}{\Alph{section}.\arabic{algocf}}

\setlength{\arrayrulewidth}{0.1mm}
\setlength{\tabcolsep}{5pt}
\renewcommand{\arraystretch}{1.5}

\section{Appendix}
\subsection{Abbreviations}
\begin{description}
\item[QG] Quantum Gravity
\item[GR] General Relativity
\item[QM] Quantum Mechanics
\item[QGIP] Quantum Gravity Information Processing
\item[S] Standard spacetime with a causally-biased structure
\item[QG] Quantum gravity spacetime with a nonfixed causal structure
\end{description}

\subsection{Notations}
The notations of the manuscript are summarized in \tref{tab2}.
\begin{center}
\begin{longtable}{||l|p{4.5in}||}
\caption{Summary of notations.}
\label{tab2}
\endfirsthead
\endhead
\hline
\textit{Notation} & \textit{Description} \\ \hline
${\rm {\mathcal Q}{\mathcal G}}$ & Quantum gravity (causally-unbiased) spacetime structure. \\ \hline 
${\rm {\mathcal S}}$ & Standard (causally-biased) spacetime structure. \\ \hline 
${\rm {\mathcal M}}$ & Measurement. \\ \hline 
$d$ & Data, $d=\left(x,\varphi _{x} ,y_{x} \right)$, where $x$ is a space-time coordination (elementary region of the space), $\varphi _{x} $ refers to the information pertinent to a choice of measurement ${\rm {\mathcal M}}$, while $y_{x} $ denotes the outcome of ${\rm {\mathcal M}}$.    \\ \hline 
$x$ & An elementary region in a space. \\ \hline 
$\varphi _{x} $ & An information pertinent to a choice of measurement ${\rm {\mathcal M}}$, for an elementary region $x$ of a space. \\ \hline 
$y_{x} $ & Outcome of measurement ${\rm {\mathcal M}}$, for an elementary region $x$ of a space. \\ \hline 
$x'$ & An elementary region in the ${\rm {\mathcal Q}{\mathcal G}}$-spacetime. \\ \hline 
$x_{A} $ & Reference region in the ${\rm {\mathcal S}}$-spacetime. \\ \hline 
$x_{B} $ & Region of interest, region of the ${\rm {\mathcal S}}$-spacetime with respect to the reference region $A$. \\ \hline 
$x'_{A} $ & Reference region in the ${\rm {\mathcal Q}{\mathcal G}}$-spacetime. \\ \hline 
$x'_{B} $ & Region of interest, region of the ${\rm {\mathcal Q}{\mathcal G}}$-spacetime with respect to the reference region $A$. \\ \hline 
$R_{x} $ & Measurement information for elementary region $x$ in the ${\rm {\mathcal S}}$-spacetime. \\ \hline 
${\rm {\mathcal O}}_{C} $ & Composite region, a set of $i$ elementary regions in the ${\rm {\mathcal S}}$-spacetime. \\ \hline 
$R_{C} $ & Measurement information for the elementary regions of composite region ${\rm {\mathcal O}}_{C} $ in the ${\rm {\mathcal S}}$-spacetime, \newline $R_{C} =\bigcup _{x\in {\rm {\mathcal O}}_{C} }R_{x}  $. \\ \hline 
$V$ & Set of all possible $R_{C} $ measurement information from all elementary regions $x$ in the ${\rm {\mathcal S}}$-spacetime, \newline $V=\bigcup _{\forall x}R_{x}  $,\newline where $R_{x} $ is the measurement information for an elementary region $x$.  \\ \hline 
$F_{x} $ & A procedure in region $x$ for the ${\rm {\mathcal S}}$-spacetime, which identifies the set of all distinct choices of $\varphi _{x} ,y_{x} $ for the elementary region $x$ as \newline $F_{x} =\bigcup _{\forall \varphi _{x} ,y_{x} }\left(x,\varphi _{x} ,y_{x} \right) $. \\ \hline 
$F_{C} $ & A procedure for composite region ${\rm {\mathcal O}}_{C} $ in the ${\rm {\mathcal S}}$-spacetime,\newline $F_{C} =\bigcup _{x\in {\rm {\mathcal O}}_{C} }F_{x}  $. \\ \hline 
$Y_{x} $ & Outcome set for a region $x$ in the ${\rm {\mathcal S}}$-spacetime, identifies the set of all distinct outcomes of a measurement ${\rm {\mathcal M}}$ for $x$, evaluated as \newline $Y_{x} =\bigcup _{\forall y_{x} }\left(x,\varphi _{x} ,y_{x} \right) $,\newline where $Y_{x} \subseteq F_{x} \subseteq R_{x} $.                                   \\ \hline 
$Y_{C} $ & Outcome set for a composite region ${\rm {\mathcal O}}_{C} $ in the ${\rm {\mathcal S}}$-spacetime, evaluated as\newline $Y_{C} =\bigcup _{x\in {\rm {\mathcal O}}_{C} }Y_{x}  $,\newline where $Y_{C} \subseteq F_{C} \subseteq R_{C} $. \\ \hline 
$R_{x'} $ & Measurement information for elementary region $x'$ in the ${\rm {\mathcal Q}{\mathcal G}}$-spacetime. \\ \hline 
${\rm {\mathcal O}}_{C'} $ & Composite region, a set of $i$ elementary regions in the ${\rm {\mathcal Q}{\mathcal G}}$-spacetime. \\ \hline 
$R_{C'} $ & Measurement information for the elementary regions of composite region ${\rm {\mathcal O}}_{C'} $ in the ${\rm {\mathcal Q}{\mathcal G}}$-spacetime, \newline $R_{C'} =\bigcup _{x'\in {\rm {\mathcal O}}_{C'} }R_{x'}  $. \\ \hline 
$V'$ & Set of all possible $R_{C'} $ measurement information from all elementary regions $x'$ in the ${\rm {\mathcal Q}{\mathcal G}}$-spacetime, \newline $V'=\bigcup _{\forall x'}R_{x'}  $,\newline where $R_{x'} $ is the measurement information for an elementary region $x'$.  \\ \hline 
$F_{x'} $ & A procedure in region $x'$ for the ${\rm {\mathcal Q}{\mathcal G}}$-spacetime, which identifies the set of all distinct choices of $\varphi _{x'} ,y_{x'} $ for the elementary region $x'$ as \newline $F_{x'} =\bigcup _{\forall \varphi _{x'} ,y_{x'} }\left(x',\varphi _{x'} ,y_{x'} \right) $. \\ \hline 
$F_{C'} $ & A procedure for composite region ${\rm {\mathcal O}}_{C'} $ in the ${\rm {\mathcal Q}{\mathcal G}}$-spacetime,\newline $F_{C'} =\bigcup _{x'\in {\rm {\mathcal O}}_{C'} }F_{x'}  $. \\ \hline 
$Y_{x'} $ & Outcome set for a region $x'$ in the ${\rm {\mathcal Q}{\mathcal G}}$-spacetime, identifies the set of all distinct outcomes of a measurement ${\rm {\mathcal M}}$ for $x'$, evaluated as \newline $Y_{x'} =\bigcup _{\forall y_{x'} }\left(x',\varphi _{x'} ,y_{x'} \right) $,\newline where $Y_{x'} \subseteq F_{x'} \subseteq R_{x'} $.                                   \\ \hline 
$Y_{C'} $ & Outcome set for a composite region ${\rm {\mathcal O}}_{C'} $ in the ${\rm {\mathcal Q}{\mathcal G}}$-spacetime, evaluated as\newline $Y_{C'} =\bigcup _{x'\in {\rm {\mathcal O}}_{C'} }Y_{x'}  $,\newline where $Y_{C'} \subseteq F_{C'} \subseteq R_{C'} $. \\ \hline 
$E$ & Event, identified via an measurement-procedure pair\newline $\begin{array}{rcl} {} & {} & {E:\left\{F_{x} ,Y_{x} \right\}} \\ {} & {=} & {\left\{\bigcup _{\forall \varphi _{x} ,y_{x} }\left(x,\varphi _{x} ,y_{x} \right) ,\bigcup _{\forall y_{x} }\left(x,\varphi _{x} ,y_{x} \right) \right\}.} \end{array}$ \\ \hline 
$A$ & Reference event, defined by the reference elementary region $x_{A} $ via the outcome-measurement pair $F_{A} ,Y_{A} $, \newline $A:\left\{F_{A} ,Y_{A} \right\}$. \\ \hline 
$B$ & Event of interest, defined via the region of interest $x_{B} $ as \newline $B:\left\{F_{B} ,Y_{B}^{i} \right\}$,\newline where $Y_{B}^{i} $ denotes a set of outcomes corresponding to $F_{B} $. \\ \hline 
$H^{{\rm {\mathcal S}}} \left(\cdot \right)$ & ${\rm {\mathcal S}}$-spacetime entropy function. \\ \hline 
$H^{{\rm {\mathcal Q}{\mathcal G}}} \left(\cdot \right)$ & ${\rm {\mathcal Q}{\mathcal G}}$-spacetime entropy function. \\ \hline 
$p_{i}^{{\rm {\mathcal S}}} \left(\cdot \right)$ & ${\rm {\mathcal S}}$-spacetime probability function,\newline $p_{i}^{{\rm {\mathcal S}}} \left(B\right)=\Pr \left(\left. Y_{B}^{i} \right|F_{B} ,D_{A} \right)$,\newline where $D_{A} $ refers to sufficient data from the past space-time region $A\in {\rm {\mathcal S}}$. \\ \hline 
$p_{i}^{{\rm {\mathcal Q}{\mathcal G}}} $ & ${\rm {\mathcal Q}{\mathcal G}}$-spacetime probability function,\newline $p_{i}^{{\rm {\mathcal Q}{\mathcal G}}} \left(B\right)=\Pr \left(\left. Y_{B}^{i} \right|Y_{A} ,F_{A} ,F_{B} \right)$. \\ \hline 
$M$  & Minkowski diagram. \\ \hline 
$E=\left(x_{E} ,ct_{E} \right)$ & Event on the Minkowski diagram, $x$ identifies the location information of events, the $ct$-axis characterizes the time information multiplied by $c$. \\ \hline 
$c$ & Constant, refers to the speed of light in the ${\rm {\mathcal Q}{\mathcal G}}$-spacetime (e.g., speed of information propagation in the ${\rm {\mathcal Q}{\mathcal G}}$-spacetime). \\ \hline 
$s^{2} $ & Lorentz parameter, $s^{2} =c^{2} t^{2} -\vec{r}^{2} ,$ where $\vec{r}^{2} =x^{2} +y^{2} +z^{2} ,$where $x,y,z,t$ the coordinates of the four-dimensional standard space-time ${\rm {\mathcal S}}$. \\ \hline 
$s_{E_{1} E_{2} }^{2} $ & $s_{E_{1} E_{2} }^{2} =c^{2} \left(t_{E_{1} } -t_{E_{2} } \right)-\left|\vec{r}_{E_{1} } -\vec{r}_{E_{2} } \right|^{2} $, for events $E_{1} =\left(\vec{r}_{E_{1} } ,ct_{E_{1} } \right)$, and $E_{2} =\left(\vec{r}_{E_{2} } ,ct_{E_{2} } \right).$ \\ \hline 
$\kappa $ & Scaling parameter, \newline $\kappa ={1\mathord{\left/ {\vphantom {1 \sqrt{\left(1-\left({\Omega \mathord{\left/ {\vphantom {\Omega  c}} \right. \kern-\nulldelimiterspace} c} \right)^{2} \right)} }} \right. \kern-\nulldelimiterspace} \sqrt{\left(1-\left({\Omega \mathord{\left/ {\vphantom {\Omega  c}} \right. \kern-\nulldelimiterspace} c} \right)^{2} \right)} } $,\newline where $\Omega =c{\textstyle\frac{c\left(t_{E_{1} } -t_{E_{2} } \right)}{x_{E_{1} } -x_{E_{2} } }} $, where $E_{1} =\left(x_{E_{1} } ,t_{E_{1} } \right)$, and $E_{2} =\left(x_{E_{2} } ,t_{E_{2} } \right).$ \\ \hline 
$\partial $ & Gradient between two events on the Minkowski diagram, $\partial ={\textstyle\frac{\Delta x}{\Delta ct}} $.  \\ \hline 
$L$ & Lightpulse, flag.  \\ \hline 
$A$, $B$ & Correlated local systems of two parties, Alice and Bob. \\ \hline 
$O_{ec} $ & Event coordinator entity in the ${\rm {\mathcal S}}$-spacetime, fixed onto the origin,\newline $O_{ec} =\left\{x_{O_{ec} } =0,ct_{O_{ec} } =0\right\}$. \\ \hline 
$O'_{ec} $ & Event coordinator entity in the ${\rm {\mathcal Q}{\mathcal G}}$-spacetime, fixed onto the origin,\newline $O'_{ec} =\left\{x'_{O_{ec} } =0,ct'_{O_{ec} } =0\right\}$. \\ \hline 
${\rm {\mathcal F}}$ & Quantum gravity function, constitutes a program ${\rm {\mathcal F}}$, as ${\rm {\mathcal F}}\left(x,n\right)={\rm i}$, where $x$ is an elementary region of the ${\rm {\mathcal Q}{\mathcal G}}$-spacetime, $n$ labels the corresponding flag beam, while ${\rm i}$ refers to the input such that for ${\rm i}=1$ a light pulse is emitted. \\ \hline 
$x$ & An elementary region in the ${\rm {\mathcal S}}$-spacetime, evaluated via  $x=\eta x'$, where $x'$ is an elementary region in the ${\rm {\mathcal Q}{\mathcal G}}$-spacetime, $\eta ={1\mathord{\left/ {\vphantom {1 \sqrt{\left(1-\beta ^{2} \right)} }} \right. \kern-\nulldelimiterspace} \sqrt{\left(1-\beta ^{2} \right)} } $, where $\beta $, $0\le \beta <1$ is the strength of the ${\rm {\mathcal Q}{\mathcal G}}$-spacetime relative to the ${\rm {\mathcal S}}$-spacetime strength $\beta _{{\rm {\mathcal S}}} =0$. \\ \hline 
$\beta $ & Strength of the ${\rm {\mathcal Q}{\mathcal G}}$-spacetime relative to the ${\rm {\mathcal S}}$-spacetime strength $\beta _{{\rm {\mathcal S}}} =0$. \\ \hline 
$\beta _{{\rm {\mathcal S}}} $ & Strength parameter of the ${\rm {\mathcal S}}$-spacetime, $\beta _{{\rm {\mathcal S}}} =0$. \\ \hline 
$\eta $ & Conversion parameter between the elementary regions of the ${\rm {\mathcal S}}$-spacetime and ${\rm {\mathcal Q}{\mathcal G}}$-spacetime,\newline $\eta ={1\mathord{\left/ {\vphantom {1 \sqrt{\left(1-\beta ^{2} \right)} }} \right. \kern-\nulldelimiterspace} \sqrt{\left(1-\beta ^{2} \right)} } $. \\ \hline 
$\theta $ & Difference of the axes $\left\{x,x'\right\}$, $\left\{ct,ct'\right\}$ of the ${\rm {\mathcal Q}{\mathcal G}}$-spacetime, and ${\rm {\mathcal S}}$-spacetime, $0^{\circ } \le \theta <45^{\circ } $, $\beta =\tan \left(\theta \right)$. \\ \hline 
$l_{scale}^{{\rm {\mathcal Q}{\mathcal G}}} $ & Unit scale of the axes of $\left\{x',ct'\right\}$ of the ${\rm {\mathcal Q}{\mathcal G}}$-spacetime, \newline $l_{scale}^{{\rm {\mathcal Q}{\mathcal G}}} =\gamma l_{scale}^{{\rm {\mathcal S}}} $,\newline where $\gamma $ is the scaling parameter evaluated as\newline $\gamma =\sqrt{{\textstyle\frac{1+\beta ^{2} }{1-\beta ^{2} }} } $,\newline and $l_{scale}^{{\rm {\mathcal S}}} $ is the unit scale of the axes of $\left\{x,ct\right\}$ of the ${\rm {\mathcal S}}$-spacetime.      \\ \hline 
$A:\left\{F_{A} ,Y_{A} \right\}$ & Measurement/outcome pair of Alice in the ${\rm {\mathcal S}}$-spacetime, \newline $F_{A} =\bigcup _{\forall \varphi _{A} ,y_{A} }\left(x_{A} ,\varphi _{x_{A} } ,y_{x_{A} } \right) $,                                         \newline $Y_{A} =\bigcup _{\forall y_{A} }\left(x_{A} ,\varphi _{x_{A} } ,y_{x_{A} } \right) $. \\ \hline 
$B:\left\{F_{B} ,Y_{B}^{i} \right\}$ & Measurement/outcome pair of Bob in the ${\rm {\mathcal S}}$-spacetime, \newline $F_{B} =\bigcup _{\forall \varphi _{B} ,y_{B} }\left(x_{B} ,\varphi _{x_{B} } ,y_{x_{B} } \right) $,                                          $Y_{B}^{i} =\bigcup _{\forall y_{B} }\left(x_{B} ,\varphi _{x_{B} } ,y_{x_{B} } \right) $. \\ \hline 
$\Delta t$ & The time difference of events $A$ and $B$ in the ${\rm {\mathcal S}}$-spacetime, evaluated as\newline $\Delta t=\eta \beta \frac{\Delta x'}{c} $,\newline where \newline $\begin{array}{rcl} {\Delta x'} & {=} & {{\textstyle\frac{1}{\eta }} \Delta x} \\ {} & {=} & {\left|x'_{B} -x'_{A} \right|} \\ {} & {=} & {\left|{\textstyle\frac{1}{\eta }} \left(x_{B} -x_{A} \right)\right|.} \end{array}$ \\ \hline 
$L_{1} $ & Flag, causally connects $A$ and $O_{ec} $ in the ${\rm {\mathcal S}}$-spacetime. \\ \hline 
$L_{2} $ & Flag, causally connects $B$ and $O_{ec} $ in the ${\rm {\mathcal S}}$-spacetime. \\ \hline 
$p_{i}^{{\rm {\mathcal Q}{\mathcal G}},\eta } $ & Probability function evaluated via the $\eta $-transformed ${\rm {\mathcal Q}{\mathcal G}}$-spacetime elementary regions,  $\eta x'_{A} =x_{A} $,$\eta x'_{B} =x_{B} $, \newline $p_{i}^{{\rm {\mathcal Q}{\mathcal G}},\eta } \left(B\right)=\Pr \left(\left. Y_{B,\eta }^{i'} \right|Y'_{A,\eta } ,F'_{A,\eta } ,F'_{B,\eta } \right)$. \\ \hline 
$f_{g}^{{\rm {\mathcal Q}{\mathcal G}},\eta } \left(\cdot \right)$ & Correlation measure functions of the ${\rm {\mathcal Q}{\mathcal G}}$-spacetime, with $\eta $-transformed ${\rm {\mathcal Q}{\mathcal G}}$-spacetime elementary regions, where $g$  identifies the correlation type. \\ \hline 
$f_{g}^{{\rm {\mathcal S}}} \left(\cdot \right)$ & Correlation measure functions of the ${\rm {\mathcal S}}$-spacetime, where $g$ identifies the correlation type. \\ \hline 
$g$ & Correlation type index. \\ \hline 
$H^{{\rm {\mathcal Q}{\mathcal G}},\eta } \left(\cdot \right)$ & The ${\rm {\mathcal Q}{\mathcal G}}$-spacetime entropy function with $\eta $-transformed arguments. \\ \hline 
$V$ & Set of all $R_{x} $, \newline $V=\bigcup _{\forall x}R_{x}  $. \\ \hline 
${\rm {\mathcal O}}_{C'} $ & A composite region in the ${\rm {\mathcal Q}{\mathcal G}}$- space. \\ \hline 
${\rm {\mathcal O}}_{C} \in {\rm {\mathcal S}}$ & A composite region in the ${\rm {\mathcal S}}$-spacetime. \\ \hline 
$I^{{\rm {\mathcal Q}{\mathcal G}},\eta } \left(\cdot \right)$ & Mutual information with $\eta $-transformed arguments of the ${\rm {\mathcal Q}{\mathcal G}}$- space. \\ \hline 
$\chi ^{{\rm {\mathcal Q}{\mathcal G}},\eta } \left(\cdot \right)$ & Holevo quantity with the $\eta $-transformed arguments of the ${\rm {\mathcal Q}{\mathcal G}}$- space. \\ \hline 
$I_{coh}^{{\rm {\mathcal Q}{\mathcal G}},\eta } \left(\cdot \right)$ & Coherent information with the $\eta $-transformed arguments of the ${\rm {\mathcal Q}{\mathcal G}}$- space. \\ \hline 
$C^{{\rm {\mathcal S}}} \left({\rm {\mathcal N}}_{AB} \right)$ & Classical capacity of quantum channel ${\rm {\mathcal N}}_{AB} $ in the ${\rm {\mathcal S}}$-spacetime. \\ \hline 
$C_{{\rm {\mathcal X}}}^{{\rm {\mathcal S}}} \left({\rm {\mathcal N}}_{AB} \right)$ & Holevo capacity of quantum channel ${\rm {\mathcal N}}_{AB} $ in the ${\rm {\mathcal S}}$-spacetime. \\ \hline 
$P^{{\rm {\mathcal S}}} \left({\rm {\mathcal N}}_{AB} \right)$ & Private classical capacity of quantum channel ${\rm {\mathcal N}}_{AB} $ in the ${\rm {\mathcal S}}$-spacetime. \\ \hline 
$Q^{{\rm {\mathcal S}}} \left({\rm {\mathcal N}}_{AB} \right)$ & Quantum capacity of quantum channel ${\rm {\mathcal N}}_{AB} $ in the ${\rm {\mathcal S}}$-spacetime. \\ \hline 
$C^{{\rm {\mathcal Q}{\mathcal G}},\eta } \left({\rm {\mathcal N}}_{AB} \right)$ & Classical capacity of ${\rm {\mathcal N}}_{AB} $ in the $\eta $-transformed ${\rm {\mathcal Q}{\mathcal G}}$-spacetime. \\ \hline 
$C_{{\rm {\mathcal X}}}^{{\rm {\mathcal Q}{\mathcal G}},\eta } \left({\rm {\mathcal N}}_{AB} \right)$ & Holevo capacity of ${\rm {\mathcal N}}_{AB} $ in the $\eta $-transformed ${\rm {\mathcal Q}{\mathcal G}}$-spacetime. \\ \hline 
$P^{{\rm {\mathcal Q}{\mathcal G}},\eta } \left({\rm {\mathcal N}}_{AB} \right)$ & Private classical capacity of ${\rm {\mathcal N}}_{AB} $ in the $\eta $-transformed ${\rm {\mathcal Q}{\mathcal G}}$-spacetime. \\ \hline 
$Q^{{\rm {\mathcal Q}{\mathcal G}},\eta } \left({\rm {\mathcal N}}_{AB} \right)$ & Quantum capacity of ${\rm {\mathcal N}}_{AB} $ in the $\eta $-transformed ${\rm {\mathcal Q}{\mathcal G}}$-spacetime. \\ \hline
\end{longtable}
\end{center}
\end{document}